\documentclass [12pt]{article}

\usepackage[dvipsnames]{xcolor}
\usepackage{hyperref}
\usepackage{graphicx}%
\usepackage{multirow}%
\usepackage{amsmath,amssymb,amsfonts}%
\usepackage{amsthm}%
\usepackage{mathrsfs}%
\usepackage[title]{appendix}%
\usepackage{xcolor}%
\usepackage{textcomp}%
\usepackage{manyfoot}%
\usepackage{booktabs}%
\usepackage{algorithm}%
\usepackage{algorithmicx}%
\usepackage{algpseudocode}%
\usepackage{listings}%

\begin{document}

\title{Polynomial potentials and nilpotent groups}
\author{
W. Schweiger, Institute of Physics\\
University of Graz, A-8010 Graz, Austria
\\ \small{\href{mailto:wolfgang.schweiger@uni-graz.at}{wolfgang.schweiger@uni-graz.at}}\\ \phantom{a}\\
W. Klink, Department of Physics and Astronomy\\
University of Iowa, Iowa City, Iowa, USA\\ \small{\href{mailto:william-klink@uiowa.edu}{william-klink@uiowa.edu}}}


\maketitle

\begin{abstract}
This paper deals with the partial solution of the energy-eigenvalue problem for one-dimensional Schr\"odinger operators of the form $H_N=X_0^2+V_N$, where $V_N=X_N^2+\alpha X_{N-1}$ is a polynomial potential of degree $(2N-2)$ and $X_i$ are the generators of an irreducible representation of a particular nilpotent group $\mathcal{G}_N$. 
Algebraization of the eigenvalue problem is achieved for eigenfunctions of the form $\sum_{k=0}^M a_k X_2^k \exp(-\int dx\, X_N)$. It is shown that the overdetermined linear system of equations for the coefficients $a_k$ has a nontrivial solution, if the parameter $\alpha$ and $(N-3)$ Casimir invariants satisfy certain constraints. This general setting works for even $N\geq 2$ and can also be applied to odd $N\geq 3$, if the potential is symmetrized by considering it as function of $|x|$ rather than $x$. It provides a unified approach to quasi-exactly solvable polynomial interactions, including the harmonic oscillator, and extends corresponding results known from the literature. Explicit expressions for energy eigenvalues and eigenfunctions are given for the quasi-exactly solvable sextic, octic and decatic potentials. The case of $E=0$ solutions for general $N$ and $M$ is also discussed.  As physical application, the movement of a charged particle in an electromagnetic field of pertinent polynomial form 
is shortly sketched.
\end{abstract}


\section{Introduction}\label{sec1}
Polynomial potentials play an eminent role in all fields of quantum physics, ranging from condensed matter to molecular, atomic, nuclear and particle physics. They comprise not only anharmonic oscillators for the description of binding forces  different from simple linear ones, but also multiwell potentials which are used to study, e.g.,  tunneling and spontaneous-symmetry-breaking phenomena. For the simplest case of a general quadratic potential, i.e. the (spatially shifted) harmonic oscillator, the energy eigenvalue problem is exactly solvable in the sense that all energy eigenvalues and corresponding eigenfunctions can be obtained in closed analytic form by algebraic means. This does not hold for potentials involving higher powers of $x$. But, surprisingly, it turned out that the energy-eigenvalue problem is at least partially solvable for the potential being a sextic polynomial, provided that the potential parameters satisfy some constraints~\cite{TURBINER1987181}. This new insight led to the notion of {\em quasi-exact solvability} which is attributed to physical models for which only a finite portion of the energy spectrum and its associated eigenfunctions can be calculated in closed analytic form by algebraic means. It triggered countless attempts to find other quasi-exactly solvable models. For a first, by far not complete, overview on different approaches and resulting quasi-exactly solvable models we refer to the nice monograph by Ushveridze~\cite{Ushveridze94}. One systematic way to construct such models rests on $sl(2,\mathbb{R})$ algebraization~\cite{Turbiner88,Kamran93}. Any one-dimensional (radial) Schr\"odinger operator which, by an appropriate change of coordinate and a gauge rotation, can be transformed into an operator that is a second-degree polynomial in the $sl(2,\mathbb{R})$ generators $J_N^+$, $J_N^0$ and $J_N^-$, is quasi-exactly solvable. A comprehensive summary of Schr\"odinger operators which admit such an $sl(2,\mathbb{R})$ algebraization can be found in Ref.~\cite{Turbiner16}. $sl(2,\mathbb{R})$ algebraization is an elegant and unifying approach to quasi-exact and even exact solvability of quantum mechanical models. However, the class of quasi-exactly solvable models is much richer, as pointed out in Ref.~\cite{Gomez-Ullate_2007}, if the notion of quasi-exact solvability is understood in a more general sense and not just restricted to models admitting $sl(2,\mathbb{R})$ algebraization.

The main objective of this paper is to find sufficient conditions for the coefficients $V_i$, under which one-dimensional Schr\"odinger operators of the form
\begin{equation}
H_{N} = - \frac{d^2}{dx^2}+\sum_{k=0}^{2N-2} V_k\, x^k \, ,\quad N\in \mathbb{N}\, ,\, N\geq 3\, ,
\end{equation}
or symmetrized versions of them,  in which the potential is considered as function of $|x|$ rather than $x$, are quasi-exactly solvable. For $N$ even and only even powers of $x$ in the potential some general results can be found in Ref.~\cite{Magyari:1981kx,Flessas:1981a}. Most of the literature on quasi-exactly solvable polynomial interactions concentrates on the sextic potential ($N=4$) containing only even powers of $x$. As a primary example for a quasi-exactly solvable potential it is often used to test new methods and approaches; see, e.g.,~\cite{Saad_2006,MAIZ2018101,Amore:2020pxg,manimegalai2020,Li:2023eji} to mention a few more recent papers. A few publications deal also with decatic ($N=6$)~\cite{Magyari:1981kx,Brandon2013,MAIZ2018101,manimegalai2020} and symmetrized quartic ($N=3$)~\cite{Skala97,Znojil16,Quesne17,Klink:2020akw} and sextic~\cite{Quesne17} potentials. 

In Ref.~\cite{Klink:2020akw} we were able to extend the class of quasi-exactly solvable symmetrized quartic potentials known from the literature~\cite{Skala97,Znojil16,Quesne17} by employing a new kind of algebraization procedure based on a nilpotent group $\mathcal{G}_3$ which we called the \lq\lq quartic group\rq\rq. We were able to show that the energy-eigenvalue problem for a Hamiltonian with symmetrized quartic potential and the structure $H_{3}=X_0^2+X_3^2+\alpha X_2$, where  $X_i$, $i=0,1,2,3$, are the generators of an irreducible representation of  $\mathcal{G}_3$, is, under two further constraints, quasi-exactly solvable. The origin of one constraint is the assumption that the energy eigenfunctions one is looking for are the product of a polynomial times an exponential factor that resembles the behavior of the ground state. For symmetrized potentials one has to demand, in addition, that the eigenfunction is continuously differentiable at $x=0$. The degree of the polynomial ansatz for the eigenfunction determines the potential parameter $\alpha$, whereas the continuity condition at $x=0$ imposes a constraint on the two Casimir invariants of $\mathcal{G}_3$. The latter means that quasi-exactly solvable quartic models with Hamiltonian 
$H_{3}=X_0^2+X_3^2+\alpha X_2$ and eigenfunctions of the assumed form can only be realized for particular irreducible representations of the quartic group.

It is now, of course, tempting to ask, whether this kind of approach can be generalized to deal with polynomial potentials of arbitrary (even) degree. First we notice that the 3-parameter Heisenberg group, which underlies the exact solvability of the harmonic oscillator, is a subgroup of the 4-parameter quartic group $\mathcal{G}_3$. It is thus suggestive to look for an $(N+1)$-parameter nilpotent group $\mathcal{G}_N$ which contains the Heisenberg group and the quartic group as subgroups and has an analogous structure. This group and its basic properties are introduced in Sec.~\ref{sec2}. Its irreducible representations, the algebra of infinitesimal generators and the Casimir invariants are shortly discussed. If one demands that the Hamiltonian has definite scaling properties under scale transformations of the position variable $x$, the possible combinations of $\mathcal{G}_N$ generators $X_i$, $i=0,1,\dots,N$, which scale like the kinetic term $X_0^2$ are restricted to $V_{N}=X_N^2+\alpha X_{N-1}$, where $\alpha$ is a free parameter and $V_{N}$ a polynomial potential of degree $(2N-2)$. In Sec.~\ref{sec3} we try to answer the question, under which conditions the Hamiltonians $H_{N}=X_0^2+V_{N}$ are quasi-exactly solvable leading to energy eigenfunctions of the form $p(x)\,\exp(-\int dx\,X_N)$, where $p(x)$ is assumed to be a polynomial of degree $M\in \mathbb{N}_0$ in the generator $X_2$. Inserting this ansatz into the Schr\"odinger equation and exploiting the algebra of generators leads to an overdetermined system of equations for the, a priori, unknown coefficients in the polynomial $p(x)$. This system is first derived and analyzed for general $N\geq 2$ and $M\in \mathbb{N}_0$. In Sec.~\ref{sec4} explicit examples for the lowest values of $M$ are worked out for unsymmetrized and symmetrized sextic,  symmetrized octic and unsymmetrized decatic ($N=4,5,6$) polynomial potentials.  These examples are compared with results from the literature. As a further application of our formalism it is also shown for arbitrary $N$ and $M=k N, k N+1\, ,\,\, k\in\mathbb{N}_0$ that the potential $V_{N,M}(x)=x^{2N-2}-(2M+N-1)\, |x|^{N-2}$ has an $E=0$ eigenvalue with the coefficients of the corresponding eigenfunction being given by a simple two-term recursion relation. Section~\ref{sec5} is devoted to a physical application. It rests on the observation that reducible representations of $\mathcal{G}_N$ give rise to a Hamiltonian which describes the movement  of a charged particle in $x$-dependent electromagnetic fields of certain polynomial form. The energy eigenfunctions of the electromagnetic field problem can thus be related to the eigenfunctions of $H_N$. In this way one ends up with quasi-exactly solvable electromagnetic field problems in three space dimensions. The main results of the paper are finally summarized in Sec.~\ref{sec6}.

%
%
\section{The nilpotent group $\mathcal{G}_N$}\label{sec2}
For the algebraization of the one-dimensional Schr\"odinger equation with polynomial interactions we consider nilpotent groups $\mathcal{G}_N$, $N\in \mathbb{N}$, $N\geq 2$, with elements
\begin{equation}
(a,b_1,b_2,\dots,b_N) \equiv (a,\vec{b}\,):=\left[\begin{array}{cc}I_{N\times N}&\vec{b}\\\vec{0}^{\,T}&1 \end{array}\right]\,\left[\begin{array}{cc}A(a)&\vec{0}\\\vec{0}^{\,T}&1 \end{array}\right],\, a,b_1,b_2,\dots,b_N \in \mathbb{R},
\end{equation}
where $I_{N\times N}$ is the $N\times N$ unit matrix and $A(a)$ an $N\times N$ matrix with elements
\begin{equation}
A_{i j}(a)=\left\{\begin{array}{ll}\frac{a^{j-i}}{(j-i)!}&\quad i\leq j\\ 0 &\quad i>j\end{array}\right.\, .
\end{equation} 
The group operation is given by
\begin{equation}
(a,\vec{b}\,)\,(a^\prime,\vec{b}^{\,\prime})=(a+a^\prime, \vec{b}+A(a) \vec{b}^{\,\prime})
\end{equation} 
and the inverse group elements are
\begin{equation}
(a,\vec{b}\,)^{-1}=(-a,-A(-a)\,\vec{b}\,)\, .
\end{equation}
By setting $b_N=0$ one ends up with a subgroup which is just the embedding of $\mathcal{G}_{N-1}$ in $\mathcal{G}_N$. Continuing by setting more and more $b_i$s zero, a whole chain of of nilpotent subgroups is generated. The Heisenberg group, which is closely connected with the harmonic oscillator~\cite{Jorgensen85,Klink94}, is, e.g., just the subgroup $\mathcal{G}_2$ and the quartic group of Ref.~\cite{Klink:2020akw} is the subgroup $\mathcal{G}_3$.

Irreducible representations of $\mathcal{G}_N$ can be obtained with the method of induced representations~\cite{Tung:1985na,Barut:1986dd}. Inducing with the Abelian subgroup
$ (0, \vec{b})\rightarrow \pi ^{\vec{\beta} }(\vec{b} ):=e^{-i \vec{\beta}\cdot \vec{b} }$, ${\vec{\beta} \in \mathbb{R}^N}$  one ends up with a unitary irreducible representation of $\mathcal{G}_N$ of the form
\begin{eqnarray}
(U^{\vec{\beta}}_{(a,\vec{b})}\phi)(x)&=&e^{-i\, \vec{\beta}^{\,T}\! A(x)\, \vec{b}}\, \phi(x+a)\, ,
\end{eqnarray}
with $(a,\vec{b}) \in \mathcal{G}_N$, $\phi \in L^2 (\mathbb{R})$.

One parameter subgroups generate representations of the Lie algebra of~$\mathcal{G}_N$:
\begin{subequations}\label{eq:generators}
\begin{align}
&(a,0,0,\dots,0) \hspace{0.5cm}\rightarrow X_0 = i\frac{\partial}{\partial x},\\
&(0, b_1,0,\dots,0) \hspace{0.4cm}\rightarrow X_1= \beta_1 ,\\
&(0,0, b_2,0,\dots,0) \, \rightarrow X_2 = \beta_2 + \beta_1 x,\\
&(0,0, 0,b_3,\dots,0)\rightarrow X_3 = \beta_3+\beta_2 x+ \frac{\beta_1 x^2}{2!},\\
& \hspace{2.7cm}\vdots\nonumber\\
&(0,0,0,\dots, 0,b_N)\rightarrow X_N  = \beta_N+\beta_{N-1} x+\dots+\frac{\beta_1 x^{N-1}}{(N-1)!},
\end{align}
\end{subequations}
with commutation relations 
\begin{equation}
[X_0, X_n]=i X_{n-1}\, , \quad n=2,3,\dots,N\, ,
\end{equation}
and all other commutators zero.

From these commutation relations one infers $(N-1)$ Casimir operators
\begin{subequations}\label{eq:casimirs}
\begin{align}
C_1 &= X_1=\beta_1,\\
C_2 &= X_1 X_3-\frac{1}{2} X_2^2=\beta_1 \beta_3-\frac{\beta_2^2}{2},\\
C_3 &= X_1^2 X_4-X_1 X_2 X_3+\frac{1}{3} X_2^3=\beta_1^2 \beta_4-\beta_1 \beta_2 \beta_3+\frac{\beta_2^3}{3}, \\
&\hspace{0.2cm}\vdots \nonumber\\
C_{k} &= \sum_{n=0}^{k-2} \frac{(-1)^n}{n!} X_1^{k-1-n} X_2^n X_{k+1-n} + \frac{(-1)^{k-1}}{(k-1)!} \frac{k-1}{k} X_2^{k}\nonumber\\
&= \sum_{n=0}^{k-2} \frac{(-1)^n}{n!} \beta_1^{k-1-n} \beta_2^n \beta_{k+1-n} + \frac{(-1)^{k-1}}{(k-1)!} \frac{k-1}{k} \beta_2^{k}\, ,\\
&\hspace{0.2cm}\vdots\nonumber
\end{align}
\end{subequations}
The irreps are then labeled by the respective values of the Casimirs. 

If one tries to solve energy-eigenvalue problems connected with $\mathcal{G}_N$, $N\geq 3$, it is most convenient to express the Lie-algebra elements $X_k$, $k\geq 3$, in terms of $X_2$ and the Casimirs. This can be done successively with the general result being
\begin{equation}\label{eq:Xreexpressed}
C_1^{k-2}X_k=\sum_{n=0}^{k-3} \frac{1}{n!} C_{k-1-n} X_2^n+\frac{1}{(k-1)!} X_2^{k-1}\, ,\quad k=3,4,\dots,N-1\, .
\end{equation}

An important restriction for the definition of our Hamiltonian will come from scaling properties of the group generators. If a unitary scaling operator is defined by
\begin{eqnarray}\label{scaling}
(S_t \phi)(x):=\sqrt{t}\, \phi( tx)\, , \quad t>0,\, \phi \in L^2 (\mathbb{R}),
\end{eqnarray}
the Lie-algebra elements scale like
\begin{eqnarray}\label{eq:scalingX}
S_t X_0 S_t^{-1}&=&t^{-1} X_0\, ,\nonumber\\
S_t X_k(\vec{\beta}) S_t^{-1}&=&t^{-(N+1-k)} X_k(\vec{\beta_t})\, , \quad k=1,2,\dots,N,\\
\hbox{with}\qquad \vec{\beta}_t&=&(t^N \beta_1,t^{N-1}\beta_2,\dots,t^2\beta_{N-1},t \beta_N)\, .\nonumber
\end{eqnarray}
%
%
%
\section{Quasi-exact solvability of Hamiltonians with polynomial interactions}\label{sec3}
Now we have all the ingredients to formulate the eigenvalue problems which we are going to investigate. We are interested in Hamiltonians with definite scaling properties that can be expressed in terms of the algebra elements $X_k$, $k=0,1,2,\dots,N$. Since the kinetic term $X_0^2$ scales like $t^{-2}$, this should also hold for the potential term. Thus one is left with Hamiltonians of the form
\begin{eqnarray}\label{eq:Hamilton}
H^{\vec{\beta}}_\alpha &:=& X_0^2 + X_N^2 +\alpha X_{N-1}\\
& =&-\frac{\partial^2}{\partial x^2} +\left(\beta_N+\beta_{N-1} x+\dots+\frac{\beta_1 x^{N-1}}{(N-1)!} \right)^2
\nonumber\\ &&+\alpha \left(\beta_{N-1} +\beta_{N-2} x +\dots+\frac{\beta_1 x^{N-2}}{(N-2)!} \right)\, ,\nonumber
\end{eqnarray}
where $\alpha\in \mathbb{R}$ is a free parameter (in addition to the $\beta$s). Looking for solutions of particular form will later on restrict the values of $\alpha$. In the simplest case, $N=2$, $H^{\vec{\beta}}_\alpha$ is just the Hamiltonian of a (spatially shifted) harmonic oscillator problem with the corresponding group $\mathcal{G}_2$ being the Heisenberg group. For $N=3,4,5,6,\dots$ one will end up with generalized quartic, sextic, octic, decatic, etc. polynomials. 

In order to solve the energy eigenvalue problem
\begin{equation}\label{eq:evproblem}
H^{\vec{\beta}}_\alpha \psi_E(x)=E \psi_E(x)
\end{equation}
we start with the ansatz
\begin{equation}\label{eq:ansatz}
\psi_E(x)=p(x)\, e^{-\int dx X_N} \, ,
\end{equation}
where it is assumed, without loss of generality, that $\beta_1>0$, such that the exponential factor vanishes for $x\rightarrow +\infty$. For \underline{$N$ even} it vanishes also in the limit $x\rightarrow -\infty$ and one ends up with normalizable solutions of the eigenvalue problem~(\ref{eq:evproblem}). For \underline{$N$ odd}, however, the exponential factor $e^{-\int dx X_N}$ diverges in the limit $x\rightarrow -\infty$. One way to obtain normalizable solutions in this case is to consider a modified problem in which the potential is symmetrized, i.e. considered as function of $|x|$ instead of $x$. For $x\geq 0$ the problem can then be solved in the usual way and one only has to continue the solution either symmetrically or antisymmetrically to $x<0$, depending on whether one wants to construct parity even or parity odd eigenfunctions. Continuous differentiability of the whole solution at $x=0$ entails a relation between the potential parameters which depends on  the considered energy eigenvalue and fixes one of the potential parameters $\beta_i$ in terms of the others. The symmetrized quartic oscillator (i.e. $N=3$), as investigated in Refs.~\cite{Skala97,Znojil16,Quesne17,Klink:2020akw}, is, e.g., an example for such symmetrized problems. One has to keep in mind, however, that potentials symmetrized in this way are non-analytic functions at $x=0$. 

\enlargethispage{10pt}

For our next steps it does not matter, whether $N$ is even or odd. For $N$ even, the ansatz (\ref{eq:ansatz}) provides already energy eigenvalues and the corresponding energy eigenfunctions we are looking for. For $N$ odd, it gives a solution of Eq.~(\ref{eq:evproblem}) which holds for $x> 0$ and has to be continued either as an even or odd function to $x<0$ in order to obtain normalizable eigenfunctions of the eigenvalue problem~(\ref{eq:evproblem}) with symmetrized potential. Inserting the ansatz~(\ref{eq:ansatz}) into Eq.~(\ref{eq:evproblem}) provides a differential equation for the function $p(x)$:
 \begin{eqnarray}\label{eq:eqp}
 -p^{\prime\prime}(x) +2 X_N\, p^{\prime}(x) +\left[(1+\alpha ) X_{N-1}- E\right]\,p(x)=0\, .
 \end{eqnarray}
Algebraization of the problem is now achieved by assuming that $p(x)$ is a polynomial function. For our purposes it turns out to be most convenient to consider it as a polynomial in the Lie-algebra element $X_2=\beta_2+\beta_1 x$ rather than $x$, i.e.
\begin{equation}\label{eq:polansatz}
p(x)=\sum_{m=0}^M a_m\, X_2^m\, ,
\end{equation}
with coefficients $a_m$ to be determined.
By inserting this polynomial form of $p(x)$ into Eq.~(\ref{eq:eqp}) and expressing $X_N$ and $X_{N-1}$ in terms of the Casimirs (cf. Eq.(\ref{eq:Xreexpressed})) one obtains
\begin{eqnarray}
&&-\sum_{m=0}^{M-2}\!(m+2) (m+1)\, C_1^2\, a_{m+2}\, X_2^m\nonumber\\&& + \sum_{n=0}^{N-3} \frac{2\, C_{N-1-n}}{n!\, C_1^{N-3}}\sum_{m=n}^{M+n-1}\!\!\!\! (m-n+1)\, a_{m-n+1}\, X_2^m\nonumber\\
&&\hspace{-0.3cm} +\sum_{m=N-1}^{M+N-2} \frac{2\, (m-N+2)}{(N-1)!\, C_1^{N-3}}\, a_{m-N+2}\, X_2^m+\sum_{n=0}^{N-4} \frac{(1+\alpha)}{n!}\frac{C_{N-2-n}}{C_1^{N-3}}\sum_{m=n}^{M+n} a_{m-n}\, X_2^m\nonumber\\
&&\hspace{-0.3cm} + \sum_{m=N-2}^{M+N-2} \frac{(1+\alpha)}{(N-2)!}\frac{1}{C_1^{N-3}}\, a_{m-N+2}\, X_2^m -\sum_{m=0}^M E\, a_m\, X_2^m =0\, .
\end{eqnarray}
Here and in the following it is understood that a sum over $n$ or $m$ should be omitted, if the upper limit is negative.
In order that this equation is satisfied for arbitrary $x$, the coefficients of $X_2^m$ have to vanish, i.e.
\begin{eqnarray}\label{eq:recursgen}
&&\hspace{-0.3cm}-(m+2) (m+1)\, C_1^2\, a_{m+2} + \sum_{n=0}^{N-3} \frac{2\, C_{N-1-n}}{n!\, C_1^{N-3}}\, (m-n+1)\, a_{m-n+1}
\nonumber\\ &&\hspace{-0.3cm}  +\sum_{n=0}^{N-4} \frac{(1+\alpha)}{n!}\frac{C_{N-2-n}}{C_1^{N-3}} a_{m-n}\, 
+ \frac{2m-N+3+\alpha (N-1)}{(N-1)!\, C_1^{N-3}} \, a_{m-N+2}\nonumber\\&&\hspace{-0.3cm}  = E\, a_m\, , \hspace{4.0cm} m=0,1,\dots,M+N-2\, .
\end{eqnarray}
To write the equations for the coefficients $a_k$ in such a general form, it is also assumed that $a_k=0$ if $k<0$ or $k>M$. 

\subsection{The harmonic oscillator}\label{sec:ho}
Let us first study  the system (\ref{eq:recursgen}) in the simplest case \underline{$N=2$}, i.e. for a (spatially shifted) \underline{harmonic oscillator}. In this case it reduces to
\begin{equation}\label{eq:recursharm}
(m+2) (m+1)\, \beta_1^2\, a_{m+2}=((2 m + 1+\alpha)\, \beta_1-E)\, a_m\, , \quad m=0,1,\dots,M\, .
\end{equation}
Written in matrix form, $\mathcal{M}\, \vec{a}=E\, \vec{a}$, this system of $(M+1)$ linear homogeneous equations for the polynomial coefficients $a_m$ could be considered as an (algebraic) eigenvalue problem which can be solved in the usual way. In view of the more general case, however, we adopt another strategy and solve it recursively starting with $m=M$. 
For $m=M$ the left-hand side of Eq.~(\ref{eq:recursharm}) vanishes and $a_M\neq 0$ implies that
\begin{equation}
E=2 \beta_1\left(\frac{1}{2}+M+\frac{\alpha}{2} \right)\, .
\end{equation}
This is just the energy of the $M$th excitation of the harmonic oscillator (apart from an energy shift by $\alpha \beta_1 $). If $M=0$, we are already done. If $M>0$, insertion of $E$ into Eq.~(\ref{eq:recursharm}), taking $m=M-1$ implies that $a_{M-1}=0$. Having $a_M\neq 0$, which fixes the normalization of the $M$th eigenfunction, and $a_{M-1}=0$, the remaining $a_m$s can be obtained from Eq.~(\ref{eq:recursharm}) by downward recursion. Apart from the normalization, the polynomial ansatz (\ref{eq:polansatz}) gives just the usual (spatially shifted) Hermite polynomial $H_M(X_2)$. By letting $M$ be any natural number $\in \mathbb{N}_0$ we are thus able to obtain the complete solution of the energy eigenvalue problem for the (shifted) harmonic oscillator. 

\subsection{General polynomial interactions}
Now let us see what happens for  \underline{$N>2$}. In this case~(\ref{eq:recursgen}) represents an overdetermined system, consisting of $(M+N-1)$ linear homogeneous equations for the coefficients $a_m$ occurring in the polynomial ansatz~(\ref{eq:polansatz}). A non-trivial solution of this homogeneous system is at most determined up to a normalization constant. Since $p(x)$ should be a polynomial of degree $M$, one can take $a_M\neq 0$ as the free coefficient which fixes the normalization. Hence one needs $M$ equations to determine the remaining $M$ coefficients $a_m$, $m=0,1,\dots,M-1$. As is shown below, the parameter $\alpha$ is fixed by setting $m=(M+N-2)$ in~(\ref{eq:recursgen}). The  coefficients $a_m$, $m<M$, can then be obtained by downward recursion, starting with $a_M$. This gives the $a_m$s, $m=0,1,\dots,M-1$, in terms of $a_M$, the energy $E$ and the $(N-1)$ Casimirs $C_j$. Having determined the $a_m$s, one is still left with $(N-2)$ equations which have to be satisfied. One of these equations serves to determine the energy eigenvalues, the remaining $(N-3)$ equations imply relations between the Casimirs which fix all the Casimirs apart from two, one of these two being $C_1$. This means that the energy-eigenvalue problem~(\ref{eq:evproblem}) with Hamiltonian~(\ref{eq:Hamilton}) admits only solutions of the form~(\ref{eq:ansatz}), with $p(x)$ given by~(\ref{eq:polansatz}), if the $(N+1)$ potential parameters $\alpha$ and $\beta_i$ satisfy certain restrictions such that one is left with a three-parameter family ($C_1=\beta_1$, $\beta_2$ and one Casimir) of potentials. 

For the highest values of $m$, i.e. $m=(M+N-2)$ and $m=(M+N-3)$, Eq.~(\ref{eq:recursgen}) reduces to
\begin{equation}\label{eq:highestcoe}
\left(\frac{2 M}{(N-1)!\, C_1^{N-3}}+\frac{(1+\alpha)}{(N-2)!\, C_1^{N-3}} \right) a_M = 0
\end{equation}
and
\begin{equation}\label{eq:nextcoe}
\left(\frac{2 (M-1)}{(N-1)!\, C_1^{N-3}}+\frac{(1+\alpha)}{(N-2)!\, C_1^{N-3}} \right) a_{M-1} = \delta_{N 3}\, E\, a_M\, ,
\end{equation}
respectively. If $p(x)$ is a polynomial of degree $M$, one has $a_M\neq 0$ and thus Eq.~(\ref{eq:highestcoe}) implies
\begin{equation}\label{eq:alphaN}
\alpha=-1-\frac{2 M}{N-1}\, .
\end{equation}
This fixes now the parameter $\alpha$ as function of the polynomial degree $M$, in contrast to the harmonic oscillator case $N=2$, where we did not have any restrictions on the potential parameters $\alpha$ and $\beta_i$.  Adopting this value of $\alpha$, Eq.~(\ref{eq:nextcoe}) can only be satisfied if
\begin{equation}\label{eq:aMm1}
a_{M-1}=- \delta_{N3}\, E\, a_M\, ,
\end{equation}
which means that $a_{M-1}=0$ if $N>3$.
This is what we can say in general about solutions of Eqs.~(\ref{eq:recursgen}) for $N\geq 3$ and $M\in\mathbb{N}_0$ arbitrary. 

\subsection{Symmetrized potentials}
As we have remarked already, the ansatz~(\ref{eq:ansatz}) provides only normalizable solutions $\psi_E(x)\in L^2(\mathbb{R})$ of the energy eigenvalue problem (\ref{eq:evproblem}) with Hamiltonian (\ref{eq:Hamilton}), if $N$ is an even number. For $N$ odd, the exponential factor diverges in the limit $x\rightarrow -\infty$. What one can do in this case is to study a  \underline{modified energy eigenvalue problem} in which the polynomial interaction is symmetrized by considering it as a function of $|x|$ rather than $x$. In our algebraic language this means that one has to deal with an energy eigenvalue problem for a \underline{modified Hamiltonian} of the form
 \begin{equation}\label{eq:hamiltonsym}
 \tilde{H}_\alpha^{\vec{\beta}}\! =\! \left\{\begin{array}{lcl} X_0^2 +X_N^2 +\alpha  X_{N-1},&&x>0\, ,\\ &&\\
\tilde{X}_0^2 +\tilde{X}_N^2 +\alpha  \tilde{X}_{N-1} ,&&x<0\, ,\\
\end{array} \right. \end{equation}
where $X_i$ and $\tilde{X}_i$ belong to different representations of the Lie algebra of $\mathcal{G}_N$, characterized by $\vec{\beta}$ and $\vec{\tilde{\beta}}$, respectively. These representations differ just in the signs of some of the $\beta$s, namely
\begin{equation}\label{eq:betatilde}
\tilde{\beta}_i=(-1)^{i+N+1}\, \beta_i\, ,\quad i=1,2,\dots,N\, .
\end{equation}
Making these sign changes when going from $x>0$ to $x<0$ is obviously equivalent to taking $|x|$ for $x\in\mathbb{R}$ and leaving the $\beta$s untouched. In this way one ends up with a potential that is an even function of $x$, i.e. $V(x)=V(-x)$, and thus one can find eigenfunctions with definite parity. 

This fact can be exploited to find also solutions of the eigenvalue problem for the symmetrized Hamiltonian~(\ref{eq:hamiltonsym}) by means of the procedure outlined above. One can see immediately that
\begin{equation}\label{eq:parity}
\tilde\psi_{E\pm}(x)=\left\{\begin{array}{lll} \phantom{\pm}(\sum_{m=0}^M a_m\, X_2^m)\, e^{-\int dx X_N}\, ,&&x>0\\
& & \\
\pm(\sum_{m=0}^M (-1)^{(N+1) m} a_m\, \tilde{X}_2^m)\, e^{-\int dx \tilde{X}_N}\, ,&&x<0\end{array}\right. \, ,
\end{equation}
is a parity even/odd (upper/lower sign) function which solves the Schr\"odinger equation 
\begin{equation}\label{eq:evproblemsym}
\tilde{H}^{\vec{\beta}}_\alpha \tilde\psi_{E\pm}(x)=E \tilde\psi_{E\pm}(x)\, ,
\end{equation}
for $x>0$ and $x<0$, if $\vec{a}$ is a solution of~(\ref{eq:recursgen}) and $E$ as well as the Casimirs are chosen such that the overdetermined system (\ref{eq:recursgen}) is satisfied. In order to be a solution of the Schr\"odinger equation on the whole real line, $\tilde\psi_{E\pm}(x)$ has to satisfy the continuity conditions
\begin{equation}\label{eq:contcond}
\lim_{\epsilon\rightarrow 0^+} \tilde\psi_{E\pm}(\epsilon)=\lim_{\epsilon\rightarrow 0^+}\tilde\psi_{E\pm}(-\epsilon)\quad\hbox{and}\quad \lim_{\epsilon\rightarrow 0^+} \tilde\psi_{E\pm}^\prime(\epsilon)=\lim_{\epsilon\rightarrow 0^+}\tilde\psi_{E\pm}^\prime(-\epsilon)\, .
\end{equation}
As one can easily check, these continuity conditions, together with the continuity of the symmetrized potential at $x=0$ guarantee that the functions $\tilde\psi_{E\pm}$ are twice continuously differentiable on the whole real axis, i.e. $\tilde\psi_{E\pm}\in C^2(\mathbb{R})$. They are thus in the domain of the self-adjoint Schr\"odinger operator associated with the differential expression $ \tilde{H}_\alpha^{\vec{\beta}}$. The domain is given by $\big\{\psi \in L^2(\mathbb{R};dx) \,|\, \psi, \psi' \in AC_\mathrm{loc}(\mathbb{R}); \, \tilde{H}_\alpha^{\vec{\beta}} \psi \in L^2(\mathbb{R}) \big\}$. Hence, $E$ is an eigenvalue  and $\tilde\psi_{E\pm}$ are eigenfunctions of the self-adjoint Schr\"odinger operator associated with $\tilde{H}^{\vec{\beta}}_\alpha$ in the strict mathematical sense (see Ref.~\cite{GNZ24}, Chap.~5).

The analysis of these continuity conditions reveals the following:\\
In the \underline{parity even case} $\tilde\psi_{E+}(x)$, as defined in Eq.~(\ref{eq:parity}), is already continuous at $x=0$. Continuity of the derivative at $x=0$ leads to the condition
\begin{equation}\label{eq:conteven}
a_0 \beta_N-\sum_{m=1}^M a_m\, (m\beta_1-\beta_2\beta_N)\, \beta_2^{m-1}=0\, .
\end{equation}
In the \underline{parity odd case} the derivative of $\tilde\psi_{E-}(x)$, as defined in Eq.~(\ref{eq:parity}), is already continuous at $x=0$. Continuity of $\tilde\psi_{E-}(x)$ at $x=0$ leads to the condition
\begin{equation}\label{eq:contodd}
\sum_{m=0}^M a_m\,\beta_2^m=0\, .
\end{equation}

As noted already, the overdetermined system of equations~(\ref{eq:recursgen}) reduces the number of independent potential parameters from $(N+1)$ to three. The continuity conditions (\ref{eq:conteven}) and (\ref{eq:contodd}) represent additional relations  between the remaining independent potential parameters. If one of these continuity conditions, either for positive or negative parity, is satisfied, one is left with a two-parameter family of (symmetrized) potentials. Note that these conditions differ for parity even and odd solutions and depend on the energy eigenvalue $E$, since the polynomial coefficients $a_m$ are functions of $E$. As a consequence one usually knows just one energy eigenvalue and the corresponding eigenfunction of either positive or negative parity for a particular set of (allowed) potential parameters. But for $M>1$ it is (sometimes) possible to choose the potential parameters such that the continuity conditions (\ref{eq:conteven}) and (\ref{eq:contodd}) are satisfied at the same time, which reduces the number of independent potential parameters further to just one, usually $C_1=\beta_1$. In this case one knows two energy eigenvalues and the corresponding parity even and odd eigenfunctions, respectively. For a detailed discussion of the symmetrized quartic oscillator (i.e. $N=3$) along the lines presented here we refer to Ref.~\cite{Klink:2020akw}.\footnote{Please notice that, for the sake of generalization, it occurred to be convenient in the present work to change some of the notations as compared to Ref.~\cite{Klink:2020akw}. This concerns in particular the numbering of the $\beta_i$s and the $X_i$s, which has been reversed.} There the class of quasi-exactly solvable symmetrized quartic oscillators known from the literature~\cite{Skala97,Znojil16,Quesne17} has been extended by employing the nilpotent \lq\lq quartic group\rq\rq \, $\mathcal{G}_3$ for algebraization of the problem.

Symmetrized problems of the form~(\ref{eq:hamiltonsym}) can, of course, also be studied for even $N$. The constraints for the continuity of parity even and odd eigenfunctions at $x=0$ are also given by Eqs. (\ref{eq:conteven}) and (\ref{eq:contodd}), respectively. The symmetrized sextic oscillator has, e.g., been investigated in Ref.~\cite{Quesne17} using the Bethe ansatz method.

%
%
 \section{Examples}\label{sec4}
In this section the general formalism developed above will be applied to various examples which correspond to particular values of $N$ and $M$. The (shifted) harmonic oscillator (i.e. $N=2$) has already been discussed in Sec.~\ref{sec:ho}. For a harmonic oscillator, symmetrized according to Eq.~(\ref{eq:hamiltonsym}), the continuity conditions~(\ref{eq:conteven}) and~(\ref{eq:contodd}) for parity even and parity odd solutions, respectively, can be satisfied at the same time, if and only if $\beta_2=0$. This means that symmetrization of the shifted harmonic oscillator potential just leads to the usual unshifted harmonic oscillator. For a detailed discussion of symmetrized quartic potentials (i.e. $N=3$) we refer to Ref.~\cite{Klink:2020akw} which contains also explicit solutions for various values of $M$.

\subsection{Sextic potential ($N=4$)}
So let us continue with the (unsymmetrized) sextic potentials. For $N=4$ the recursion relation~(\ref{eq:recursgen}) reduces to
\begin{eqnarray}\label{eq:recurssextic}
&& -6(m+2) (m+1)\, C_1^3\, a_{m+2}+12 (m+1)\, C_3\, a_{m+1}\nonumber\\ &&+6\, \left[\,(\alpha+2 m+1)\, C_2-C_1 E \right] a_m\nonumber\\
&&+(3\alpha+2 m-1)\, a_{m-2}=0
\, , \qquad m=0,1,\dots,M+2\, .
\end{eqnarray}
The recursion relation~(\ref{eq:recurssextic}) for $m=M+2, M+1$ implies (cf. Eqs.~(\ref{eq:alphaN}) and~(\ref{eq:aMm1}))
\begin{equation}\label{eq:alaM1}
\alpha=-\frac{3+2 M}{3}\qquad\hbox{and}\qquad a_{M-1}=0\, .
\end{equation}
In the following we will give explicit solutions for the lowest values of $M$.
\medskip

\noindent\underline{M=0} ($\alpha=-1$):\\
Equation~(\ref{eq:recurssextic}) for $m=0$ implies that
\begin{equation}
C_1\, E\,  a_0=0\, .
\end{equation}
A non-trivial solution of the energy eigenvalue problem is thus only achieved when $E=0$. The corresponding eigenfunction is
\begin{equation}\label{eq:sextic0}
\Psi_0^\mathrm{sext}(x)=a_0\,e^{-\int dx\,X_4}=a_0\,e^{-\beta_4 x-\frac{\beta_3}{2} x^2-\frac{\beta_2}{6} x^3-\frac{\beta_1}{24} x^4}\, ,
\end{equation}
with $a_0$ an appropriate normalization constant. This means that a sextic potential of the form $V^\mathrm{sext}_{0}=X_4^2-X_3$ has a zero energy ground state with the corresponding eigenfunction given by Eq.~(\ref{eq:sextic0}). Apart from $\beta_1>0$ there are no further restrictions on the potential parameters $\beta_i$. Hence the sextic potential $V^\mathrm{sext}_{0}$ is not necessarily spatially symmetric, unless it is assumed that $\beta_2=\beta_4=0$, as is usually done in the literature~\cite{TURBINER1987181,Saad_2006,Turbiner16,MAIZ2018101,Amore:2020pxg,manimegalai2020,Li:2023eji}. In this case $\Psi_0^\mathrm{sext}(x)$ is a parity-even $E=0$ energy eigenfuction.

\medskip

\noindent\underline{M=1} ($\alpha=-\frac{5}{3}$):\\
From Eq.~(\ref{eq:alaM1}) we infer that $a_0=0$. Setting $m=0,1$ in Eq.~(\ref{eq:recurssextic}) leads to the two equations
\begin{eqnarray}
m=1:&&    \left[4\,C_2-3\, C_1\, E \right] a_1=0\, , \nonumber\\
m=0:&&    C_3\, a_1=0\, . 
\end{eqnarray}
These two equations imply that 
\begin{equation}\label{eq:sextic1}
\Psi_1^\mathrm{sext}(x)=a_1\,X_2\,e^{-\int dx\,X_4}
\end{equation}
solves the Schr\"odinger equation with sextic potential $V^\mathrm{sext}_{1}=X_4^2-\frac{5}{3} X_3$, if
\begin{equation}\label{eq:esextic1}
E=\frac{4}{3} \frac{C_2}{C_1} \quad\hbox{and}\quad C_3=\beta_1^2 \beta_4-\beta_1 \beta_2 \beta_3+\frac{\beta_2^3}{3}=0\, .
\end{equation}
The wave function $\Psi_1^\mathrm{sext}(x)$ has one node and corresponds thus to a first excited state. Note that $C_3=0$ does not necessarily mean that $\beta_2=\beta_4=0$ and hence a spatially symmetric potential. But if this is the case,  $\Psi_1^\mathrm{sext}(x)$ is a parity-odd energy eigenfuction.

\medskip
\noindent\underline{M=2} ($\alpha=-\frac{7}{3}$):\\
According to Eq.~(\ref{eq:alaM1}) one now has $a_1=0$. Setting $m=0,1,2$ in Eq.~(\ref{eq:recurssextic}) leads to the three equations
\begin{eqnarray}
m=2:&&    \left[\,8\, C_2-3\, C_1\, E\, \right] \,a_2-2\, a_0=0\, , \nonumber\\
m=1:&&   C_3\, a_2=0\, , \nonumber\\
m=0:&&    -6\, C_1^3\, a_2- \left[\,4\, C_2+3\, C_1\, E\, \right]\, a_0=0\, . 
\end{eqnarray}
The $m=1$ relation implies again that $C_3=0$. From the remaining two equations one has
\begin{equation}
a_0=\left[4\, C_2-\frac{3}{2}\,C_1\, E_\mp\right] a_2
\end{equation}
with 
\begin{equation}
E_\mp=\frac{2}{3}\frac{C_2}{C_1}\mp 2 \sqrt{\frac{C_2^2}{C_1^2}+\frac{1}{3} C_1}\, .
\end{equation}
For the sextic potential $V^\mathrm{sext}_{2}=X_4^2-\frac{7}{3} X_3$ (with $C_3=0$) we thus know two energy eigenvalues with the corresponding eigenfunctions
\begin{equation}\label{eq:sextic2}
\Psi^\mathrm{sext}_{2\,\mp}(x)=a_2\,\left[X_2^2+\left(4\, C_2-\frac{3}{2}\,C_1\, E_\mp\right)\right]\,e^{-\int dx\,X_4}
\end{equation}
describing a ground state (no node) and a second excited state (two nodes).\footnote{Note that a polynomial with even/odd degree and real coefficients has an even/odd number of real zeroes.}

\medskip
\noindent\underline{M=3} ($\alpha=-3$):\\
According to Eq.~(\ref{eq:alaM1}) one now has $a_2=0$. Setting $m=0,1,2,3$ in Eq.~(\ref{eq:recurssextic}) leads to the four equations
\begin{eqnarray}
m=3:&&    -3 \left[\,4\, C_2- C_1\, E\, \right] \,a_3+2\, a_1=0\, , \nonumber\\
m=2:&&   -6\,C_3\,a_3+a_0=0\, , \nonumber\\
m=1:&&    6\, C_1^2\, a_3+E\, a_1=0\, ,\nonumber\\
m=0:&&    -2\, C_3\, a_1+\,\left[\,2\, C_2+C_1\, E\, \right]\, a_0=0\, . 
\end{eqnarray}
The $m=3$ and $m=1$ relations hold if
\begin{equation}
a_1=-6\,\frac{C_1^2}{E_\mp} \,a_3\qquad\hbox{with}\qquad 
E_\mp=\frac{2 C_2}{C_1}\mp 2 \sqrt{\frac{C_2^2}{C_1^2}+C_1}\, .
\end{equation}
With these results it follows that the remaining two equations are satisfies if and only if $C_3=0$ (and hence $a_0=0$). For the sextic potential $V^\mathrm{sext}_{3}=X_4^2-3 X_3$ (with $C_3=0$) we thus know two energy eigenvalues with the corresponding eigenfunctions
\begin{equation}\label{eq:sextic3}
\Psi^\mathrm{sext}_{3\,\mp}(x)=a_3\,\left[X_2^3-6\,\frac{C_1^2}{E_\mp} X_2 \right] \,e^{-\int dx\,X_4}
\end{equation}
describing a first excited state (one node) and a third excited state (three nodes).

\medskip

It can be checked that $C_3=0$ is also a sufficient condition for the solvability of Eqs.~(\ref{eq:recurssextic}) if $M>3$. $C_3=0$ guarantees that all the coefficients $a_m$, $m$ odd/even, vanish, if M is an even/odd number. The equations for the vanishing coefficients are then satisfied automatically. The remaining equations, apart from one, fix the non-vanishing coefficients (apart from the normalization $a_M$). The last equation is a kind of compatibility condition for the non-vanishing coefficients and allows to determine the energy eigenvalues $E$.  

Assuming $C_3=0$, one ends up with a cubic equation for the energy eigenvalues in case of $M=4$ and $M=5$. Shortly summarized, the results for these two cases are:

\medskip
\noindent\underline{M=4} ($\alpha=-\frac{11}{3}$):\\
The eigenenergies $E$ of the the ground state, the second and the fourth excited states are solutions of the cubic equation
\begin{eqnarray}\label{eq:ev4}
&&\left( E + \frac{8}{3} \frac{C_2}{C_1} \right) \left( E - \frac{4}{3} \frac{C_2}{C_1} \right) \left( E - \frac{16}{3} \frac{C_2}{C_1} \right) \nonumber\\&&- 8 C_1 \left( E + \frac{8}{3} \frac{C_2}{C_1} \right)-\frac{8}{3} C_1 \left( E - \frac{16}{3} \frac{C_2}{C_1} \right)=0\, .
\end{eqnarray}
The coefficients determining the corresponding eigenfunctions are
\begin{eqnarray}
a_2&=& -\frac{3}{2} C_1 \left( E - \frac{16}{3} \frac{C_2}{C_1} \right) a_4\, , \nonumber\\
a_0&=&\frac{9}{8} C_1^2 \left[ \left( E - \frac{16}{3} \frac{C_2}{C_1} \right) \left( E - \frac{4}{3} \frac{C_2}{C_1} \right) - 8 C_1\right] a_4\, ,\quad a_3=a_1=0\, ,
\end{eqnarray}
with $a_4$ an appropriate normalization constant.

\medskip
\noindent\underline{M=5} ($\alpha=-\frac{13}{3}$):\\
In this case the eigenenergies $E$ of the the first, third and fifth excited states are solutions of the cubic equation
\begin{eqnarray}\label{eq:ev5}
&&\left( E + \frac{4}{3} \frac{C_2}{C_1} \right) \left( E - \frac{8}{3} \frac{C_2}{C_1} \right) \left( E - \frac{20}{3} \frac{C_2}{C_1} \right)\nonumber\\ && - \frac{40}{3} C_1 \left( E + \frac{4}{3} \frac{C_2}{C_1} \right)-8 C_1 \left( E - \frac{20}{3} \frac{C_2}{C_1} \right)=0\, .
\end{eqnarray}
The coefficients determining the corresponding eigenfunctions are
\begin{eqnarray}
a_3&=& -\frac{3}{2} C_1 \left( E - \frac{20}{3} \frac{C_2}{C_1} \right) a_5\, ,\\
a_1&=&\frac{9}{8} C_1^2 \left[ \left( E - \frac{20}{3} \frac{C_2}{C_1} \right) \left( E - \frac{8}{3} \frac{C_2}{C_1} \right) - \frac{40}{3} C_1\right] a_5,\, a_4=a_2=a_0=0\, ,\nonumber
\end{eqnarray}
with $a_5$ an appropriate normalization constant.

\begin{figure}[t!]
\begin{centering}
\includegraphics[width=0.49\textwidth]{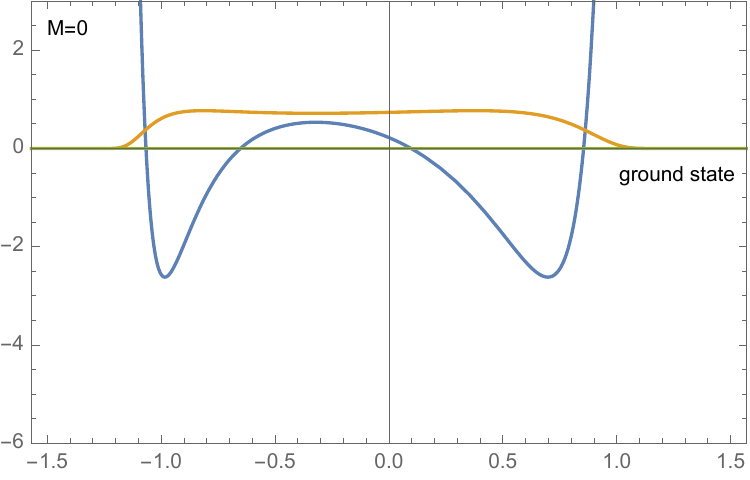}\hfill
\includegraphics[width=0.49\textwidth]{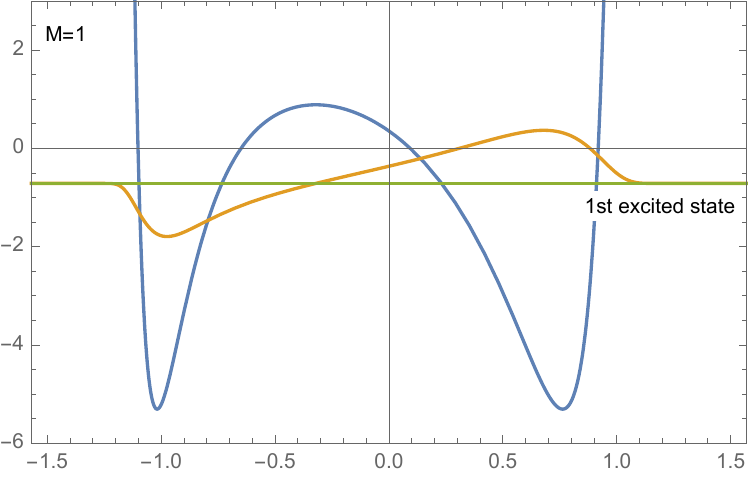}
\par\end{centering}\vspace{-0.3cm}
\caption{The sextic potential $(X_4^2+\alpha X_3)$ for $\alpha=-1$ (left) and $\alpha=-\frac{5}{3}$ (right), $\beta_1=6.$, $\beta_2=2.$, $\beta_3=-0.2$ and $\beta_4=\frac{\beta_2 \beta_3}{\beta_1}-\frac{\beta_2^3}{3 \beta_1^2}$ (i.e. $C_3=0$) along with the corresponding analytically calculable energy eigenvalues and eigenfunctions. Potential and wave functions are plotted as functions of $y=\arctan x$. The normalization of the wave function has been chosen such that $\int_{-\pi/2}^{\pi/2} dy\,{\Psi_M^\mathrm{sext}}^2(x(y))=1$.}\label{fig:sext01}
\end{figure}


\begin{figure}[t!] 
\begin{centering}
\includegraphics[width=0.49\textwidth]{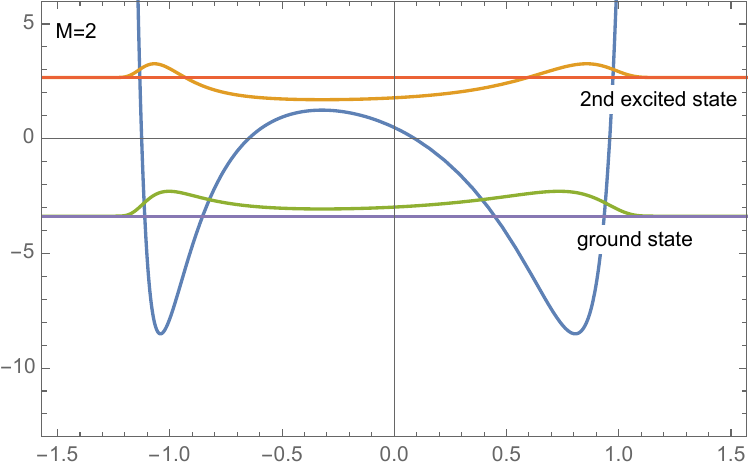}\hfill
\includegraphics[width=0.49\textwidth]{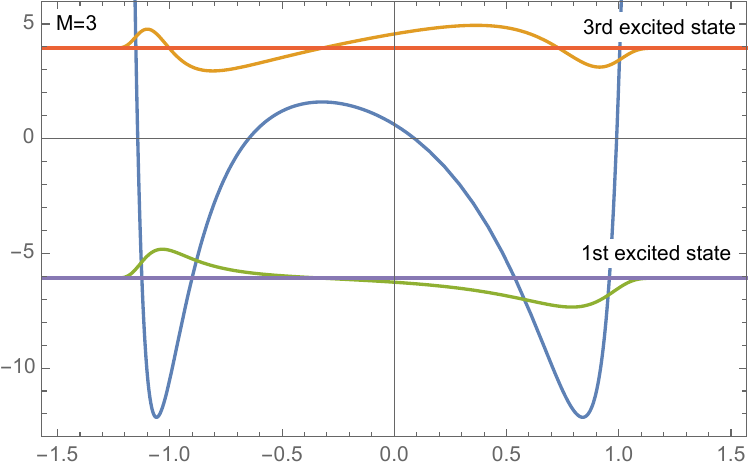}
\par\end{centering}\vspace{-0.3cm}
\caption{Same as in Fig.~\ref{fig:sext01}, but for $\alpha=-\frac{7}{3}$ (left) and $\alpha=-3$ (right), corresponding to $M=2$ and $M=3$, respectively.}\label{fig:sext23}\vspace{-0.3cm}
\end{figure}

\begin{figure}[h!]
\begin{centering}
\includegraphics[width=0.49\textwidth]{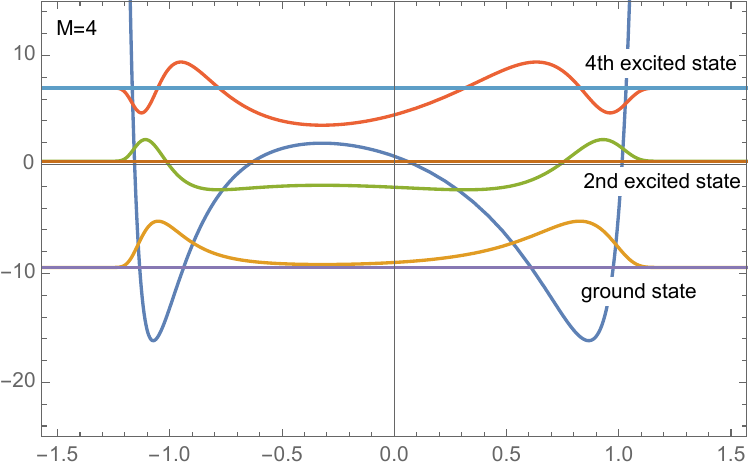}\hfill
\includegraphics[width=0.49\textwidth]{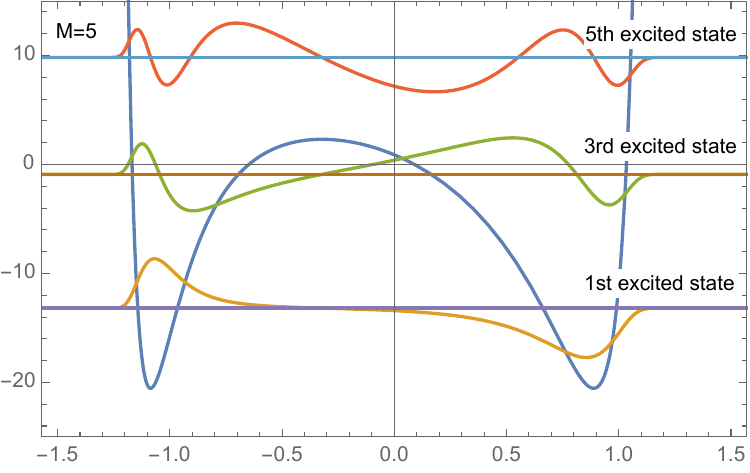}
\par\end{centering}
\caption{Same as in Fig.~\ref{fig:sext01}, but for $\alpha=-\frac{11}{3}$ (left) and $\alpha=-\frac{13}{3}$ (right), corresponding to $M=4$ and $M=5$, respectively. For better visability the wave functions have been normalized to $10$ rather than~$1$.} \label{fig:sext45}
\end{figure}

Examples of quasi-exactly solvable sextic potentials along with the analytically calculable eigenvalues and eigenfunctions for $M=0,1,2,3,4,5$ are plotted in Figs.~\ref{fig:sext01}--\ref{fig:sext45}. The potential parameters for these examples were chosen to be $\beta_1=6.$, $\beta_2=2.$, $\beta_3=-0.2$ and $\beta_4=\frac{\beta_2 \beta_3}{\beta_1}-\frac{\beta_2^3}{3 \beta_1^2}$ (such that $C_3=0$)  which leads to double-well potentials. Note that the parameter $\alpha$ becomes increasingly negative with increasing~$M$. 

\noindent\underline{Remark:}
The class of quasi-exactly solvable sextic potentials $V=X_4^2+\alpha X_3$, restricted by the solvability condition $C_3=0$, does not only include double-well potentials, but also anharmonic oscillator potentials with just one minimum or even triple-well potentials with three minima and two (relative) maxima. It is related to the usual spatially symmetric sextic oscillators known from the literature~\cite{TURBINER1987181,Saad_2006,Turbiner16,MAIZ2018101,Amore:2020pxg,manimegalai2020,Li:2023eji} by means of a translation.\footnote{Our quasi-exactly solvable sextic oscillators (restricted by $C_3=0$) are, apart of a constant, obtained from those in Ref.~\cite{Turbiner16} by applying a translation $x\rightarrow x+\frac{\beta_2}{\beta_1}$ and setting $(4n+2k)=-3 (\alpha+1)$, $a=\frac{\beta_1}{6}$, $b=\frac{2 \beta_1 \beta_3-\beta_2^2}{2 \beta_1}$, where $k=0,1$, depending on whether $M$ is even or odd, respectively.} 

\medskip

\noindent\underline{Remark:} Analytic solutions of the energy-eigenvalue equations~(\ref{eq:ev4}) and~(\ref{eq:ev5}) are easily obtained in the special case $C_2=0$.

\medskip

\noindent\underline{Remark:} Whereas $C_3=0$ is a necessary and sufficient condition for the solvability of Eqs.~(\ref{eq:recurssextic}) in the case of $M=0,1,2,3$, it is only a sufficient condition for $M>3$. As one can check, e.g., for $M=4$, Eqs.~(\ref{eq:recurssextic}) can also be solved for $C_3=\pm(-\frac{27}{14} C_2 (C_1^3+\frac{16}{49} C_2^2))^{1/2}$. In both cases one has only one eigenvalue, $E=\frac{40\, C_2}{21\, C_1}$, with corresponding eigenfunction of the form~(\ref{eq:ansatz}) and $p(x)$ given by~(\ref{eq:polansatz}). For $M=5$ there are even 4 ways to choose $C_3$ different from zero. Unlike the case $C_3=0$, our quasi-exactly solvable potentials for $C_3\neq 0$ cannot be related to the usual spatially symmetric sextic oscillators by means of a translation. 

\medskip

\noindent\underline{Remark:}  For $C_3=0$ the $(\tilde{M}+1)$-dimensional subspace of the Hilbert space consisting of functions of the form $\sum_{m=0}^{\tilde{M}} a_m X_2^{2m+k} \mathrm{exp}(-\int dx X_4)$, $a_m\in\mathbb{R}$ and $k=0,1$, is invariant under the action of the sextic Schr\"odinger operator
$X_0^2+X_4^2-(3+2k+4 \tilde{M})\, X_3/3$. This explains, why one can find $(\tilde{M}+1)$ energy eigenvalues and it also explains the equivalence of our approach and $sl(2,\mathbb{R})$ algebraization (if one assumes $C_3=0$). For $C_3\neq 0$, however, the finite dimensional subspace of functions of the form~(\ref{eq:ansatz}), with $p(x)$ being a polynomial of degree $M>3$, is not an invariant subspace of the sextic Schr\"odinger operator. It nevertheless may contain single eigenfunctions of the sextic Schr\"odinger operator, if $C_3$ satisfies an appropriate constraint. 
Our general notion of quasi-exact solvability allows for such single solutions which are not accessible by means of the usual $sl(2,\mathbb{R})$ algebraization. For (symmetrized) quartic, octic, decatic, or even higher-order polynomial potentials such finite dimensional invariant subspaces (with dimension $>1$) would not even exist.

\subsection{Symmetrized sextic potential ($N=4$)}
Let us next consider the energy eigenvalue problem for the symmetrized sextic potential, i.e. for the Hamiltonian given by Eq.~(\ref{eq:hamiltonsym}) with $N=4$. From the foregoing discussion we know already that $C_3=0$ guarantees the solvability of Eqs.~(\ref{eq:recurssextic}) and implies that the coefficients $a_m$ must be zero for odd/even index $m$, if $M$ is even/odd.   As with the symmetrized quartic oscillator, continuous differentiability of the energy eigenfunctions at $x=0$ leads to the additional constraints (\ref{eq:conteven}) and (\ref{eq:contodd}) for the potential parameters $\beta_i$, depending on whether one wants parity even or odd eigenfunctions. In the following we will give explicit solutions for the lowest values of M.

\medskip

\noindent\underline{M=0} ($\alpha=-1$):\\
\noindent There is no non-trivial \underline{parity odd} solution for $M=0$.\\
The \underline{parity even} solution in this case has to satisfy the constraint (cf. Eq.~(\ref{eq:conteven}))
\begin{equation}
a_0\, \beta_4=0 \qquad \hbox{which implies}\qquad \beta_4=0\, .
\end{equation}
This means that 
\begin{equation}
\Psi_0^{\mathrm{sext}+}(x)=a_0\, e^{-(\beta_3\frac{x^2}{2}+\beta_2 \frac{|x|^3}{6}+\beta_1 \frac{x^4}{24})}
\end{equation}
is an $E=0$ eigenfunction of the symmetrized sextic potential, provided that $\beta_4=0$. Setting $\beta_1=6$, $\beta_2=6a$ and $\beta_3=-2b$ we reproduce (apart from a shift in energy) the result given in Eqs.~(24) and (25) of Ref.~\cite{Quesne17}.

\medskip

\noindent\underline{M=1} ($\alpha=-\frac{5}{3}$):\\
The solvability condition $C_3=0$ and the continuity condition at $x=0$ fix two of the four $\beta$s in terms of the remaining ones.\\ In the \underline{parity odd} case, $C_3=0$ and Eq.~(\ref{eq:contodd}) are satisfied if and only if 
\begin{equation}
\beta_2=\beta_4=0\, .
\end{equation}
The analytically calculable energy eigenvalue of the resulting potential is given by (see Eq.~(\ref{eq:esextic1}))
\begin{equation}
E=\frac{4 \beta_3}{3} 
\end{equation}
and the corresponding eigenfunction is
\begin{equation}
\Psi_1^{\mathrm{sext}-}(x)=a_1\,x\, e^{-(\beta_3\frac{x^2}{2}+\beta_1 \frac{x^4}{24})}\, .
\end{equation}
With one node it is the wave function of a first excited state. Since $\beta_2=\beta_4=0$, the \lq\lq symmetrized\rq\rq\ potential as well as the eigenfunction $\Psi_1^{\mathrm{sext}-}(x)$ are analytic at $x=0$.  It is just a special case ($\beta_2=\beta_4=0$) of Eq.~(\ref{eq:sextic1}) and setting $\beta_1=6$ it agrees (apart from a shift in energy) with the well known (analytic) sextic oscillator of Ref.~\cite{Turbiner16}. With $\beta_1=6$ and $\beta_3=-2 b$  it is also a special case ($a=0$) of the symmetrized potentials given in Eq.~(26) of Ref.~\cite{Quesne17}.\\

\noindent In the \underline{parity even} case, the condition $C_3=0$ and Eq.~(\ref{eq:conteven}) are most easily solved for $\beta_2$ and $\beta_3$ with the result
\begin{equation}\label{eq:M1evensym}
\beta_2=\frac{\beta_1}{\beta_4}\quad\hbox{and}\quad \beta_3=\frac{\beta_1}{3 \beta_4^2}+\beta_4^2\, .
\end{equation}
The analytically calculable energy eigenvalue of the resulting potential is given by
\begin{equation}
E=-\frac{2\, \beta_1}{9\, \beta_4^2}+\frac{4}{3}\beta_4^2\, .
\end{equation}
The corresponding eigenfunction is (see Eq.~(\ref{eq:sextic1}))
\begin{equation}\label{eq:N1sym}
\Psi_1^{\mathrm{sext}+}(x)=a_1\left( \beta_2+\beta_1 |x|\right) \, 
e^{-(\beta_4 |x|+\beta_3\frac{x^2}{2}+\beta_2 \frac{|x|^3}{6}+\beta_1 \frac{x^4}{24})}
\end{equation}
with $\beta_2$ and $\beta_3$ given above. Depending on the sign of $\beta_4$, it is either a ground-state wave function without nodes, or the wave function of a second excited state with two nodes. Our class of quasi-exactly solvable symmetrized sextic potentials which give rise to energy eigenfunctions of the form~(\ref{eq:N1sym}) differs again from the one given in Ref.~\cite{Quesne17}. Only under the restriction that $a$ and $b$ are related by $a^4+a^2 b+\frac{1}{2}=0$ (which means that $c=1/a$), the potential given in in Eq.~(26) of Ref.~\cite{Quesne17}  is recovered (apart from a constant term) by setting $\beta_1=6$, $\beta_2=6 a$, $\beta_3=-2 b$, $\beta_4=\frac{1}{a}$. Under these circumstances we are also able to reproduce the energy eigenvalue and the corresponding eigenfunction given in Eq.~(27) of Ref.~~\cite{Quesne17}.

\medskip

\noindent\underline{M=2} ($\alpha=-\frac{7}{3}$):\\
In the \underline{parity odd} case, $C_3=0$ and Eq.~(\ref{eq:contodd}) are most easily solved for $\beta_1$ and $\beta_3$ with the result
\begin{equation}\label{eq:M2odd}
\beta_1=2 \beta_2 \beta_4 \qquad \hbox{and}\qquad \beta_3=\frac{\beta_2}{6 \beta_4}+2\beta_4^2\, .
\end{equation} 
The analytically calculable energy eigenvalue for the resulting potential is
\begin{equation}
E=\frac{\beta_2}{9\beta_4}+\frac{16}{3} \beta_4^2\, .
\end{equation}
The corresponding eigenfunction is given by Eq.~(\ref{eq:sextic2}) for $x>0$ and has to be continued antisymmetrically to $x<0$.\\

\noindent In the \underline{parity even} case, the two conditions $C_3=0$ and Eq.~(\ref{eq:conteven}) allow for two solutions
\begin{eqnarray}\label{eq:M2evensym}
\beta_{1\pm}&=&\frac{2 \beta_2^2+3 \beta_2\beta_4^3\pm \sqrt{4 \beta_2^4-18\beta_2^3 \beta_4^3+9 \beta_2^2\beta_4^6}}{15 \beta_4^2}\, ,\nonumber\\
\beta_{3\pm}&=&\frac{14 \beta_2^2+21 \beta_2\beta_4^3\mp 3 \sqrt{4 \beta_2^4-18\beta_2^3 \beta_4^3+9 \beta_2^2\beta_4^6}}{30 \beta_2\beta_4}\, .
\end{eqnarray} 
The corresponding analytic solutions for the eigenenergies are
\begin{equation}
E_\pm=\frac{- 34 \beta_2^2+39 \beta_2\beta_4^3\pm 3 \sqrt{4 \beta_2^4-18\beta_2^3 \beta_4^3+9 \beta_2^2\beta_4^6}}{45 \beta_2 \beta_4}\, .
\end{equation}
These equations have to be understood in such a way that one should take either the upper or the lower sign. The corresponding eigenfunctions are given by Eq.~(\ref{eq:sextic2}) for $x>0$ and have to be continued symmetrically to $x<0$. In addition, the free parameters $\beta_2$ and $\beta_4$ are restricted by the requirement that $\beta_{1}$ and $\beta_3$ should be real numbers and $\beta_{1}>0$.

\begin{figure}[t!]
\begin{centering}
\includegraphics[width=0.49\textwidth]{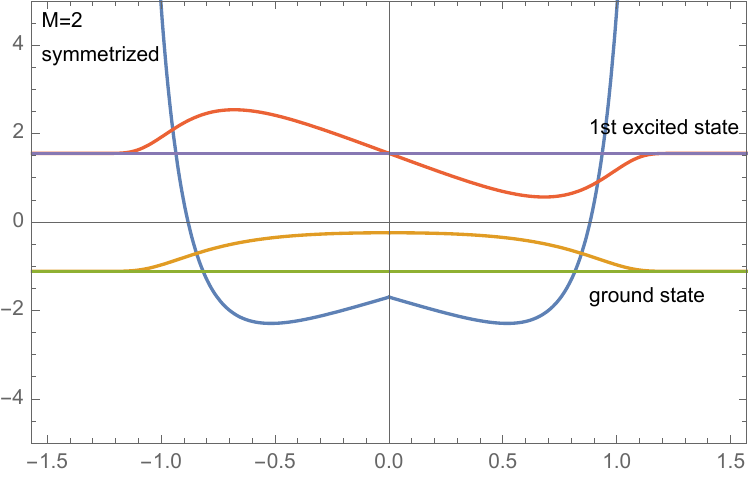}\hfill
\includegraphics[width=0.49\textwidth]{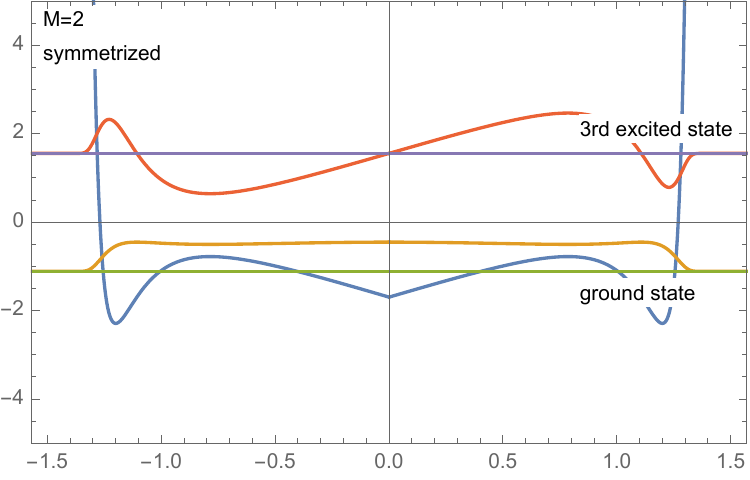}\hfill
\par\end{centering}
\caption{The symmetrized sextic potential for $\alpha=-\frac{7}{3}$, $\beta_4=0.5$ (left) and $\beta_4=-0.5$ (right), $\beta_1=16\beta_4^4$, $\beta_2=8\beta_4^3$ and $\beta_3=\frac{10}{3} \beta_4^2$ along with the corresponding analytically calculable energy eigenvalues and eigenfunctions of the parity even ground state and the parity odd excited state. Potential and wave functions are plotted as functions of $y=\arctan x$. The normalization of the wave function has been chosen such that $\int_{-\pi/2}^{\pi/2} dy\,{\Psi_2^{\mathrm{sext}\pm}}^2(x(y))=1$.}\label{fig:sext2sym}
\end{figure}

Interestingly, the expressions for $\beta_{1+}$ and $\beta_{3+}$, given in (\ref{eq:M2evensym}), can be made equal to those for $\beta_1$ and $\beta_3$ given in~(\ref{eq:M2odd}), if one takes
\begin{equation}
\beta_2=8\beta_4^3 \quad\hbox{such that} \quad \beta_1=16 \beta_4^4\quad\hbox{and}\quad\beta_3=\frac{10}{3}\beta_4^2\, .
\end{equation}
This means that one now has a one-parameter family of symmetrized sextic potentials for which one knows two energy eigenvalues,
\begin{equation}
E_{\mathrm{even}}=-\frac{40}{9} \beta_4^2 \quad\hbox{and}\quad E_{\mathrm{odd}}=\frac{56}{9} \beta_4^2\, ,
\end{equation}
corresponding to a parity even and a parity odd energy eigenstate, respectively. The particular choice $\beta_4=0.5$, e.g., leads to a double-well potential for which energies and eigenfunctions of the ground state and the first excited state can be calculated analytically. The outcome is shown on the left-hand side of Fig.~\ref{fig:sext2sym}. For $\beta_4=-0.5$ one would rather obtain a triple-well potential and the analytically calculable eigenenergies and eigenfunctions are those of the ground state and the third excited state (see Fig.~\ref{fig:sext2sym} right).

Again, our class of symmetrized sextic potentials giving rise to positive or negative parity energy eigenfunctions of the form (\ref{eq:sextic2}) (for $x>0$) does not coincide with the one in Ref.~\cite{Quesne17}. Whereas the symmetrized sextic oscillators given in Ref.~\cite{Quesne17} comprise the usual analytic sextic oscillators of Ref.~\cite{Turbiner16} as special cases (parameters $a=c=0$), this is not the case for our quasi-exactly solvable symmetrized sextic oscillators. To obtain the usual analytic sextic oscillators of Ref.~\cite{Turbiner16}, we should set $\beta_2=\beta_4=0$. But according to the constraints Eqs.~(\ref{eq:M2odd}) and (\ref{eq:M2evensym}) this would lead to a vanishing potential. 

In the parity-odd case the potential given in Eq.~(28) of Ref.~\cite{Quesne17}  can be recovered (apart from a constant) by setting $\beta_1=6$, $\beta_2=6 a$, $\beta_3=-2 b$, $\beta_4=\frac{1}{2 a}$, provided that $a$ and $b$ are related by $4 a^4+4 a^2 b+1=0$ (which means that $c=1/(2a)$). In the parity-even case the potential given in Eq.~(30) of Ref.~\cite{Quesne17}  can be recovered (apart from a constant) by setting $\beta_{1-}=6$, $\beta_2=6 a$, $\beta_{3-}=-2 b$, $\beta_4=c$, provided that $a$, $b$, $c$, $x_1$ and $x_2$ are related by $8 a^2+2 a c^3-3 c^2+4 a b c=0$, $8 a^2+2 a c^3-5 c^2-4 a^3 c=0$\footnote{These two constraints are obtained by replacing the $\beta_i$s in the \lq\lq -\rq\rq\  solution (\ref{eq:M2evensym}) by their expressions in terms of $a$, $b$ and $c$. As one can check, they follow also from Eq.~(32) of Ref.~\cite{Quesne17}, if $x_1+x_2=2 a$.} and $x_1+x_2=2 a$. Under these circumstances we are also able to reproduce Eqs.~(29) and (31) of Ref.~\cite{Quesne17}.

\subsection{Symmetrized octic potential ($N=5$)}
For $N=5$, only the symmetrized version of the potential $X_5^2+\alpha X_4$ is quasi-exactly solvable by means of the ansatz~(\ref{eq:parity}). For $N=5$ the recursion relation~(\ref{eq:recursgen}) reduces to
\begin{eqnarray}\label{eq:recursoctic}
&&\hspace{-1.0cm} -12(m+2) (m+1) C_1^4 a_{m+2}+24 (m+1) C_4 a_{m+1}\nonumber\\&&\hspace{-1.0cm}
+12 \left[\,(\alpha+2 m+1) C_3-C_1^2 E \right] a_m\nonumber\\
&&\hspace{-1.0cm}+12 (\alpha+m) C_2 a_{m-1}+(2 \alpha+m-1) a_{m-3}=0
\, ,\quad m=0,1,\dots,M+3\, .
\end{eqnarray}
The recursion relation~(\ref{eq:recursoctic}) for $m=M+3, M+2$ implies (cf. Eqs.~(\ref{eq:alphaN}) and~(\ref{eq:aMm1}))
\begin{equation}\label{eq:alaM2}
\alpha=-1-\frac{M}{2}\qquad\hbox{and}\qquad a_{M-1}=0\, .
\end{equation}
In the following we will give explicit solutions for the lowest values of $M$.
\medskip

\noindent\underline{M=0} ($\alpha=-1$):\\
\noindent There is no non-trivial \underline{parity odd} solution for $M=0$.\\
Continuity of the \underline{parity even} solution is guaranteed if (cf. Eq.~(\ref{eq:conteven}))
\begin{equation}
a_0\, \beta_5=0 \qquad \hbox{and hence}\qquad \beta_5=0\, .
\end{equation}
This means that 
\begin{equation}
\Psi_0^{\mathrm{oct}+}(x)=a_0\, e^{-(\beta_4\frac{x^2}{2}+\beta_3 \frac{|x|^3}{6}+\beta_2 \frac{x^4}{24}+\beta_1 \frac{|x|^5}{120})}
\end{equation}
is an $E=0$ eigenfunction of the symmetrized octic oscillator, provided that $\beta_5=0$. 

\medskip

\noindent\underline{M=1} ($\alpha=-\frac{3}{2}$):\\
Setting $m=3,4$ in Eq.~(\ref{eq:recursoctic}) implies that $a_0=0$ and $\alpha=-\frac{3}{2}$, respectively. Equation~(\ref{eq:recursoctic}) for $m=0,1,2$ leads to the solvability conditions
\begin{equation}\label{eq:contoct1}
C_2=C_4=0
\end{equation}
and the energy eigenvalue
\begin{equation}\label{eq:EocticM1}
E=\frac{3C_3}{2C_1^2}\, .
\end{equation}
The solvability conditions (\ref{eq:contoct1}) and the continuity condition at $x=0$ fix three of the five $\beta$s in terms of the remaining ones.\\ In the \underline{parity odd} case, $C_2=C_4=0$ and Eq.~(\ref{eq:contodd}) are satisfied if and only if 
\begin{equation}
\beta_2=\beta_3=\beta_5=0\, .
\end{equation}

\medskip

\noindent In the \underline{parity even} case, the condition $C_2=C_4=0$ and Eq.~(\ref{eq:conteven}) are easily solved for $\beta_2$, $\beta_3$ and $\beta_4$ with the result
\begin{equation}
\beta_2=\frac{\beta_1}{\beta_5}\, ,\quad\beta_3=\frac{\beta_1}{2\beta_5^2}\, ,\quad\beta_4=\frac{\beta_1}{8\beta_5^3} +\beta_5^2\, .
\end{equation}
The parity even and parity odd eigenfunctions corresponding to the eigenenergy (\ref{eq:EocticM1}) (with the respective restrictions on the $\beta$s) are given by Eq.~(\ref{eq:parity}) with $a_0=0$ and $a_1$ an appropriate normalization constant.

\medskip

\noindent\underline{M=2} ($\alpha=-2$):\\
Setting $m=3,4,5$ in Eq.~(\ref{eq:recursoctic}) implies that 
\begin{equation}\label{eq:coeffM2octic}
a_0=6 C_2 a_2\, ,\qquad a_1=0
\end{equation}
and $\alpha=-2$, respectively. Equation~(\ref{eq:recursoctic}) for $m=0,1,2$ leads to the solvability conditions
\begin{equation}\label{eq:contoct2}
C_3=-\frac{C_1^4}{12 C_2}\, ,\qquad C_4=\frac{3 C_2^2}{2}
\end{equation}
and the energy eigenvalue
\begin{equation}\label{eq:EocticM2}
E=-\frac{C_1^2}{4 C_2}\, .
\end{equation}
In the \underline{parity odd} case, the solvability conditions~(\ref{eq:contoct2}) and the continuity condition~(\ref{eq:contodd}) are most easily solved for $\beta_1$, $\beta_3$ and $\beta_4$ with the result
\begin{equation}\label{eq:octM2odd}
\beta_1=2 \beta_2\,\beta_5\, ,\quad\beta_3=\frac{\beta_2}{6 \beta_5}\, ,\quad\beta_4=2 \beta_5^2\, .
\end{equation} 

\medskip

\begin{figure}[t!]
\begin{centering}
\includegraphics[width=0.49\textwidth]{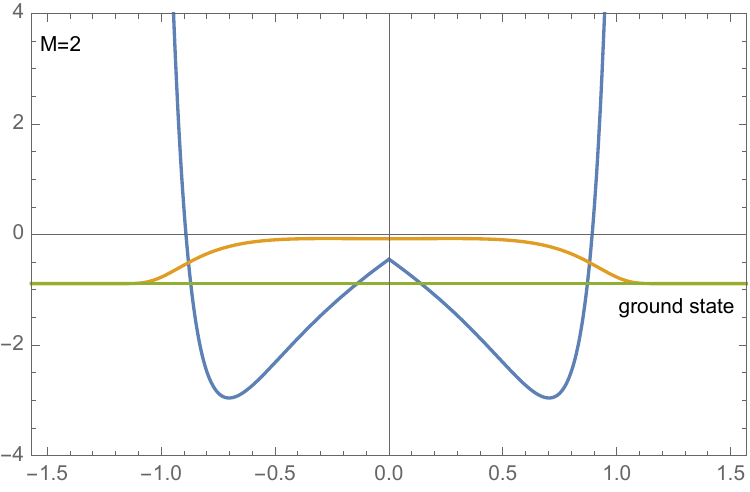}\hfill
\includegraphics[width=0.49\textwidth]{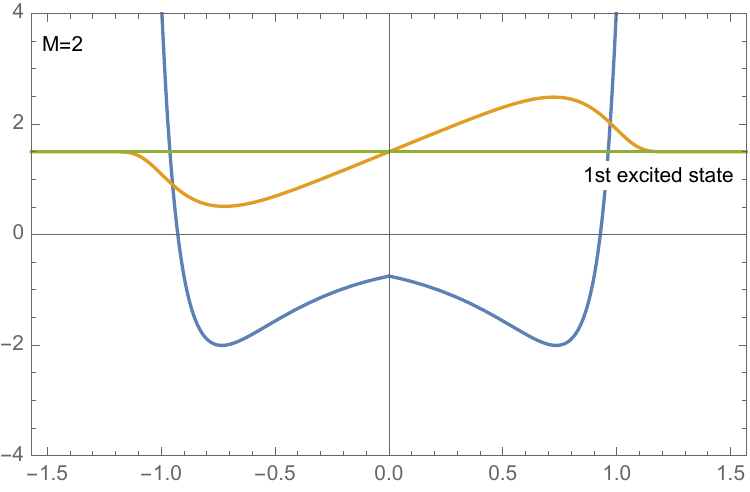}\hfill
\par\end{centering}
\caption{The symmetrized octic potential for $\alpha=-2$, $\beta_2=2.$, $\beta_5=0.5$ and $\beta_1$, $\beta_2$, $\beta_4$ chosen according to Eq.~(\ref{eq:octM2even}), upper sign (left), and Eq.~(\ref{eq:octM2odd}) (right), respectively. Drawn are also the corresponding analytically calculable energy eigenvalues and eigenfunctions of even (left) and odd parity (right). Potential and wave functions are plotted as functions of $y=\arctan x$. The normalization of the wave function has been chosen such that $\int_{-\pi/2}^{\pi/2} dy\,{\Psi_2^{\mathrm{oct}\pm}}^2(x(y))=1$.}\label{fig:oct2sym}
\vspace{0.5cm}
\end{figure}

\noindent In the \underline{parity even} case, the the solvability conditions~(\ref{eq:contoct2}) and the continuity condition~(\ref{eq:conteven}) allow for two solutions
\begin{eqnarray}\label{eq:octM2even}
\beta_{1\pm}&=&\frac{\beta_2^2+3 \beta_2\beta_5^4\pm \sqrt{\beta_2^4-9\beta_2^3 \beta_5^4+9 \beta_2^2\beta_5^8}}{15 \beta_5^3}\, ,\nonumber\\
\beta_{3\pm}&=&\frac{2 \beta_2^2+3 \beta_2\beta_5^4\mp \sqrt{\beta_2^4-9\beta_2^3 \beta_5^4+9 \beta_2^2\beta_5^8}}{3 \beta_2 \beta_5}\, ,\nonumber\\
\beta_{4\pm}&=&\frac{7 \beta_2^2+36 \beta_2\beta_5^4\mp 8 \sqrt{\beta_2^4-9\beta_2^3 \beta_5^4+9 \beta_2^2\beta_5^8}}{30 \beta_2 \beta_5^2}\, .
\end{eqnarray} 
Again, these equations have to be understood in such a way that one should take either the upper or the lower sign. The free parameters $\beta_2$ and $\beta_5$ are restricted by the requirement that $\beta_{1}$, $\beta_3$ and $\beta_4$ are real numbers and $\beta_{1}>0$.
The parity even and parity odd eigenfunctions corresponding to the eigenenergy (\ref{eq:EocticM2}) (with the respective restrictions on the $\beta$s) are given by Eqs.~(\ref{eq:parity}) and (\ref{eq:coeffM2octic}) with $a_2$ an appropriate normalization constant. Unlike the case of the symmetrized quartic or sextic potential, there is no common set of potential parameters for which both, a parity even and a parity odd solution of the form~(\ref{eq:parity}), can be found. Figure~\ref{fig:oct2sym} shows two examples of octic double well potentials for which one eigenvalue with corresponding polynomial eigenfunction of either odd or even parity can be calculated analytically. In the literature we were only able to find numeric or perturbative treatments of (unsymmetrized) octic anharmonic oscillators~\cite{Ivanov_1998,jafarpour2001,Pathak_2002,Mahapatra:2004ch}.

\vspace{0.5cm}

\subsection{Decatic potential ($N=6$)}
Like the (unsymmetrized) sextic potential, the decatic potential 
\begin{equation}\label{eq:Vdec}
V^\mathrm{dec}(x)=X_6^2+\alpha X_5
\end{equation} 
is quasi-exactly solvable by means of the ansatz (see Eqs.~(\ref{eq:ansatz}) and (\ref{eq:polansatz}))
\begin{equation}\label{eq:ansdec}
\Psi_M^\mathrm{dec}(x)=\sum_{m=0}^M a_m\,X_2^m\,e^{-\int dx X_6}\, ,
\end{equation}
provided that the potential parameters $\alpha$ and $\beta_i$, $i=1,2,\dots,5$, satisfy 4 constraints, which guarantee that the overdetermined system of equations (cf. Eq.~(\ref{eq:recursgen}))

\vfill\break

\begin{eqnarray}\label{eq:recursdecic}
&&\hspace{-0.3cm} -120 (m+2) (m+1) C_1^5 a_{m+2}\! +\! 240 (m+1) C_5 a_{m+1}\!\nonumber\\&&\hspace{-0.3cm} +\! 120 \left[\,(\alpha+2 m+1) C_4-C_1^3 E \right] a_m\nonumber\\
&&\hspace{-0.3cm}+120 (\alpha+m) C_3 a_{m-1}+20 (3 \alpha+2 m-1) C_2 a_{m-2}+(5 \alpha+2 m-3) a_{m-4}=0\, ,
\nonumber\\ && \phantom{a}\hspace{6.0cm} m=0,1,\dots,M+4\, .
\end{eqnarray}
can be solved for the coefficients $a_i$, $i=0,1,\dots,M-1$ ($a_M$ serves as wave function normalization) and the energy eigenvalue $E$. Taking $N=6$, Eqs.~(\ref{eq:alphaN}) and (\ref{eq:aMm1}) imply that
\begin{equation}\label{eq:alaM2dec}
\alpha=-1-\frac{ 2M}{5}\qquad\hbox{and}\qquad a_{M-1}=0\, .
\end{equation}
In the following we will shortly summarize the quasi-exactly solvable decatic potentials for $M=0,1,2,3,4,5$. \medskip

\noindent\underline{M=0} ($\alpha=-1$):\\
In this special case there are, apart from $\beta_1>0$, no further restrictions on the potential parameters $\beta_i$. The decatic potential $V^\mathrm{dec}_{0}=X_6^2-X_5$ has an
\begin{equation}
E=0
\end{equation}
ground state with the corresponding eigenfunction of the form~(\ref{eq:ansdec}). For $M=0$ the class of quasi exactly solvable potentials given in Ref.~\cite{Brandon2013} is just a subset of $V^\mathrm{dec}_{0}=X_6^2-X_5$. It contains only spatially symmetric potentials (for which $\beta_2=\beta_4=\beta_6=0$). We reproduce Eq.~(30) of Ref.~\cite{Brandon2013} (apart from the constant term in $V^\mathrm{dec}_{0}(x)$), if we set $\beta_1=120$, $\beta_3=-3$, $\beta_5=\frac{3}{8}$ and $\beta_2=\beta_4=\beta_6=0$.

\medskip

\noindent\underline{M=1} ($\alpha=-\frac{7}{5}$):\\
The potential parameters have to satisfy the constraints
\begin{equation}
C_2=C_3=C_5=0\, ,
\end{equation}
which are, e.g., satisfied by choosing
\begin{equation}\label{eq:constdec1}
\beta_3=\frac{\beta_2^2}{2\beta_1}\, ,\quad \beta_4=\frac{\beta_2^3}{6\beta_1^2}\, ,\quad \beta_6=-\frac{\beta_2^5}{30\beta_1^4}+\frac{\beta_2 \beta_5}{\beta_1}\, .
\end{equation}
The analytically calculable energy
\begin{equation}
E=\frac{8 C_4}{5 C_1^3}
\end{equation}
and wave function of the form~(\ref{eq:ansdec}) with
\begin{equation}
a_0=0
\end{equation}
and $a_1$ an appropriate normalization constant correspond to a first excited state. For $M=1$ the class of quasi exactly solvable potentials given in Ref.~\cite{Brandon2013} is not completely covered by our approach. In order to end up with a spatially symmetric potential, one has to take $\beta_2=0$ and hence, according to Eq.~(\ref{eq:constdec1}), also $\beta_4=\beta_6=0$. But this means that the $x^8$ term in $V^\mathrm{dec}_{1}=X_6^2-\frac{7}{5}X_5$ vanishes, so that only potentials of Ref.~\cite{Brandon2013} with vanishing parameter $b$ ($b=0$) are included in our approach. Under this restriction we are able to reproduce the energy eigenvalue (apart from the constant term in $V_1^\mathrm{dec}$) and the constraints on the potential parameters given in Eq.~(29) of Ref.~\cite{Brandon2013} by setting  $a=\frac{\beta_1^2}{14400}$, $b=0$, $c=\frac{\beta_1 \beta_5}{60}$ and $\beta_2=\beta_4=\beta_6=0$.

\medskip

\noindent\underline{M=2} ($\alpha=-\frac{9}{5}$):\\
The potential parameters have to satisfy the constraints
\begin{equation}
C_3=C_5=0\, ,\quad C_4=-\frac{C_1^5}{16 C_2}+\frac{8}{10} C_2^2\, ,
\end{equation}
which can be uniquely solved in still reasonable simple form for $\beta_4$, $\beta_5$ and $\beta_6$. The analytically calculable energy eigenvalue and the corresponding eigenfunction of the form~(\ref{eq:ansdec}) are given by
\begin{equation}
E=-\frac{C_1^2}{5 C_2}-\frac{16 C_2^2}{25 C_1^3}\quad \hbox{and}\quad  a_0= 8 C_2 a_2\, , \, a_1=0\, ,
\end{equation}
with $a_2$ an appropriate normalization constant.
\medskip

\noindent\underline{M=3} ($\alpha=-\frac{11}{5}$):\\
If the potential parameters satisfy the constraints
\begin{equation}\label{eq:const103}
C_3=C_5=0\, ,\quad C_4=-\frac{C_1^5}{8 C_2}+\frac{4}{5} C_2^2\, ,
\end{equation}
which can be uniquely solved for $\beta_4$, $\beta_5$ and $\beta_6$, the analytically calculable energy eigenvalue and the corresponding eigenfunction of the form~(\ref{eq:ansdec}) are given by
\begin{equation}
E=-\frac{3 C_1^2}{5 C_2}+\frac{16 C_2^2}{25 C_1^3}\quad \hbox{and}\quad  a_0=a_2=0\, ,\, a_1= 12 C_2 a_3\, ,
\end{equation}
with $a_3$ an appropriate normalization constant. In addition to the set of constraints~(\ref{eq:const103}) one can find 2 other sets of constraints, with $C_3$ and $C_5$ different from zero, which provide also quasi-exactly solvable decatic potentials. 
\medskip

\noindent\underline{M=4} ($\alpha=-\frac{13}{5}$):\\
If the potential parameters satisfy one of the 2 sets of constraints (either upper or lower sign)
\begin{equation}\label{eq:const104}
C_3=C_5=0\, ,\, C_4=\frac{320 C_2^4-135 C_1^5 C_2\mp\sqrt{2025C_1^{10} C_2^2-1920 C_1^5 C_2^5+4096
   C_2^8}}{480 C_2^2}\, ,
\end{equation}
which can be uniquely solved for $\beta_4$, $\beta_5$ and $\beta_6$, the analytically calculable energy eigenvalue and the corresponding eigenfunction of the form~(\ref{eq:ansdec}) are given by
\begin{eqnarray}
&&\hspace{-0.7cm}E=\frac{-15 C_1^5 C_2\pm\sqrt{2025C_1^{10} C_2^2-1920 C_1^5 C_2^5+4096
   C_2^8}}{50 C_1^3 C_2^2}\quad \hbox{and} \quad a_1=a_3=0\, ,\nonumber\\
&&   \hspace{-0.7cm}a_0=\frac{64 C_2^4-45 C_1^5 C_2 \mp\sqrt{2025 C_1^{10}C_2^2-1920 C_1^5
  C_2^5+4096 C_2^8}}{2
   C_2^2} a_4\, ,\, a_2= 16 C_2 a_4\, , \nonumber\\&&
\end{eqnarray}
with $a_4$ an appropriate normalization constant. In addition to the two sets of constraints~(\ref{eq:const104}) one can find 4 other sets of constraints, with $C_3$ and $C_5$ different from zero, which provide also quasi-exactly solvable decatic potentials. 
\medskip

\begin{figure}[t!]
\begin{centering}
\includegraphics[width=0.49\textwidth]{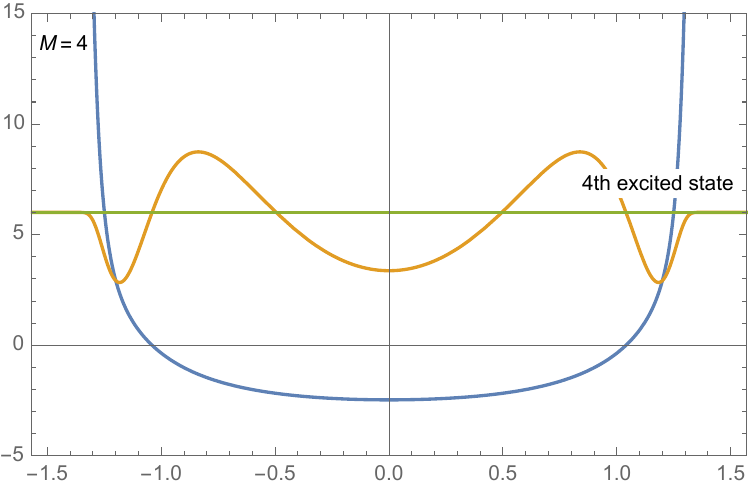}\hfill
\includegraphics[width=0.49\textwidth]{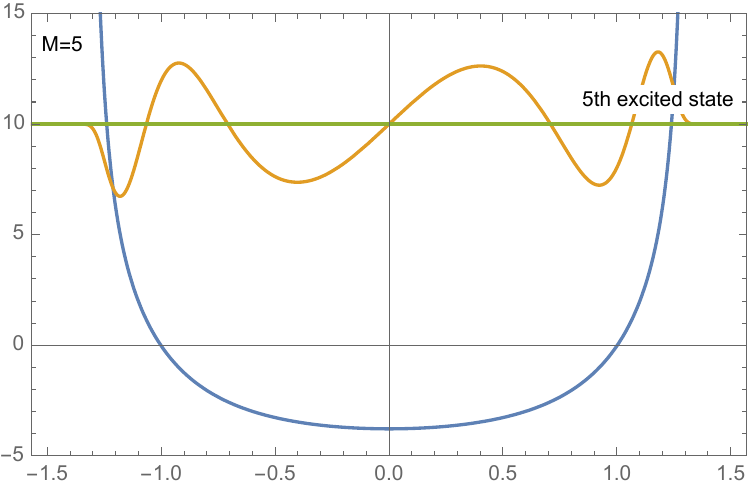}\hfill
\par\end{centering}
\caption{The decatic potential for $\alpha=-\frac{13}{5}$, $\beta_1=0.5$, $\beta_3=-0.1$, $\beta_5=0.949938$, $\beta_2=\beta_4=\beta_6=0$,  (left) and $\alpha=-3$, $\beta_1=0.5$, $\beta_3=-0.1$, $\beta_5=1.26$, $\beta_2=\beta_4=\beta_6=0$ (right),  along with the corresponding analytically calculable energy eigenvalues and eigenfunctions. Potential and wave functions are plotted as functions of $y=\arctan x$. For better visability the normalization of the wave function has been chosen such that $\int_{-\pi/2}^{\pi/2} dy\,\Psi_{4(5)}^2(x(y))=10$. The parameters $\beta_4$, $\beta_5$ and $\beta_6$ are chosen such that they satisfy the respective constraints~(\ref{eq:const104}) (upper sign) or~(\ref{eq:const105}).
}\label{fig:dec45}
\end{figure}

\noindent\underline{M=5} ($\alpha=-3$):\\
The potential parameters have to satisfy the constraints
\begin{equation}\label{eq:const105}
C_3=C_5=0\, ,\quad C_4= - \frac{C_1^5}{4 C_2}+\frac{1}{2} C_2^2\, ,
\end{equation}
which can be uniquely solved in still reasonable simple form for $\beta_4$, $\beta_5$ and $\beta_6$. The analytically calculable energy eigenvalue and the corresponding eigenfunction of the form~(\ref{eq:ansdec}) are given by
\begin{equation}
E=-\frac{2 C_1^2}{C_2}\quad \hbox{and}\quad  a_0=a_2=a_4=0\, ,\, a_1=60 C_2^2 a_5\, ,  a_3=20 C_2 a_5\, ,
\end{equation}
with $a_5$ an appropriate normalization constant. Surprisingly, unlike the cases $M=3,4$ there is only one set of constraints which leads to quasi-exactly solvable decatic potentials. For $M=6$, however, one has again several sets of constraints for $C_3$, $C_4$ and $C_5$ which lead to quasi-exactly solvable decatic potentials.
\medskip

Examples of quasi-exactly solvable decatic potentials along with the analytically calculable eigenvalues and eigenfunctions for $M=4,5$ are plotted in Fig.~\ref{fig:dec45}. The eigenvalues and eigenfunctions in these particular cases correspond to a fourth and fifth excited state, respectively.

\medskip

\noindent\underline{Remark:} As one can check, $C_3=C_5=0$ is a sufficient condition for the solvability of Eqs.~(\ref{eq:recursdecic}), even for arbitrary $M$. Assuming that $C_3=C_5=0$, Eq.~(\ref{eq:recursdecic}) reduces to a four-term recursion relation which connects every second $a_k$. Setting $m=M, M-1$ fixes $\alpha$ and $a_{M-1}=0$, respectively. By downward recursion, setting $m=M+2,M+1,M,\dots,4$, the coefficients $a_{M-2}, a_{M-4},\dots$ can be expressed in terms of (the normalization) $a_M$ and, as a consequence of $a_{M-1}=0$, one obtains $a_{M-3}=a_{M-5}=\dots=0$. Two of the four remaining equations for $m=0,1,2,3$ are then satisfied identically and the other two equations fix $C_4$ and the energy eigenvalue $E$.

\subsection{$E=0$ solutions for arbitrary $N$ and $M$}
With increasing $N$ and $M$ it becomes obviously more and more complicated to find analytic solutions of Eqs.~(\ref{eq:recursgen}). For $N>2$ non-trivial solutions for the coefficients $a_i$ are only obtained, if one takes $\alpha=-1-\frac{2M}{N-1}$ and the Casimirs $C_i$ satisfy $N-3$ constraints. This means that only $3$ of the $N$ potential parameters $\beta_i$ can be chosen freely. If the potential is symmetrized (for $N$ odd), one of the continuity conditions at $x=0$, Eq.~(\ref{eq:conteven}) or Eq.~(\ref{eq:contodd}), poses a further constraint on the $\beta_i$s which reduces the number of free potential parameters to just $2$. A tremendous simplification of Eqs.~(\ref{eq:recursgen}) is, however, achieved, if one assumes that 
\begin{equation}\label{eq:Cizero}
C_2=C_3=\dots=C_{N-1}=0
\end{equation}
and concentrates on \underline{$E=0$} solutions. Under these circumstances Eqs.~(\ref{eq:recursgen}) reduce to the two-term recursion relation
\begin{equation}\label{eq:recursspec}
a_{m-N}=-\frac{(N-1)!}{2}\frac{m (m-1)}{N+M-m} C_1^{N-1} a_m\, ,\qquad m=0,1,\dots,M+N-3\, .
\end{equation}
Here we have already used that $\alpha=-1-\frac{2M}{N-1}$. Setting $m=2,3,\dots,N-1$ this recursion relation implies (with $a_{m<0}=0$) that $a_2=a_3=\dots=a_{N-1}=0$. $a_0$ and $a_1$, on the other hand, can be different from zero. But this means that non-trivial solutions of Eq.~(\ref{eq:recursspec}) with $a_M\neq 0$ (which serves as normalization) are only obtained if
\begin{equation}
M=N k\,  ,\, N k+1\, ,\quad k\in \mathbb{N}_0\, .
\end{equation}
By downward recursion, Eq.~(\ref{eq:recursspec}) provides then $a_{M-N}, a_{M-2N}\,,\dots$, starting with $a_M\neq 0$. All other $a_i$s vanish. The restrictions~(\ref{eq:Cizero}) imply that
\begin{equation}
\beta_i=\frac{\beta_2^{i-1}}{(i-1)! \,\beta_1^{i-2}}\, ,\qquad i=3,4,\dots,N\, ,
\end{equation}
so that only $\beta_1$ and $\beta_2$ can be chosen freely. If the potential is symmetrized (which is necessary for odd $N$), continuity of the solution at $x=0$ relates $\beta_1$ and $\beta_2$ so that one is left with one open parameter.

For the presentation of explicit examples we simplify the problem further and assume that
\begin{equation}
\beta_2=0\qquad \hbox{and}\qquad \beta_1=(N-1)!\, .
\end{equation}
As a consequence one has $\beta_3=\beta_4=\dots=\beta_N=0$ and the (symmetrized) potential $X_N^2+\alpha X_{N-1}$ reduces to
\begin{equation}\label{eq:potNM}
V_{N,M}(x)=x^{2N-2}-(2M+N-1)\, |x|^{N-2}\, , \qquad N\geq 2\, ,\, \,M=k N, k N+1\, ,\,\, k\in\mathbb{N}_0 .
\end{equation}
Note that this potential agrees with its unsymmetrized version, if $N$ is even. Furthermore, the continuity conditions Eq.~(\ref{eq:conteven}) and Eq.~(\ref{eq:contodd}) are satisfied automatically, if all the $\beta_i$s, apart from $\beta_1$, vanish. For $M=0,1$ the potentials $V_{N,M}(x)$ are special cases ($m=0$, $g=1$, $n=N-1$) of potentials considered in Ref.~\cite{Turbiner:1979hb}.

\begin{figure}[t!]
\begin{centering}
\includegraphics[width=0.49\textwidth]{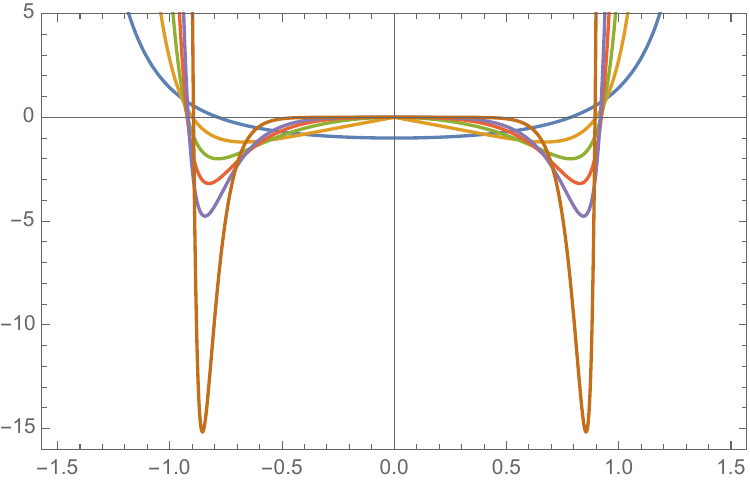}\hfill
\includegraphics[width=0.49\textwidth]{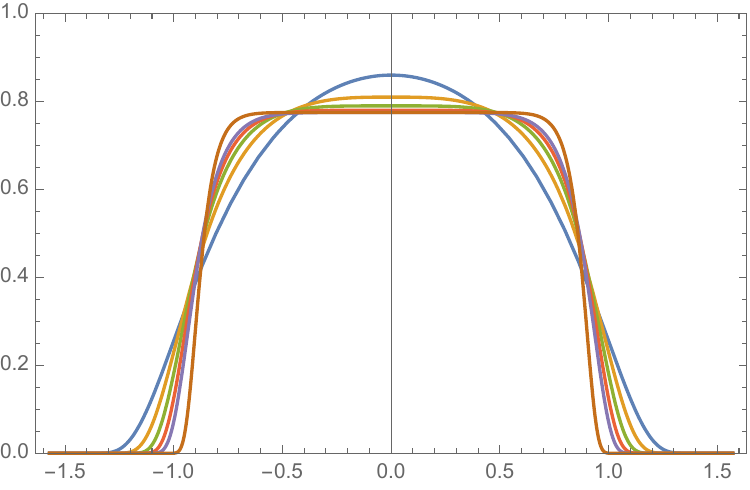}\hfill\\
\includegraphics[width=0.49\textwidth]{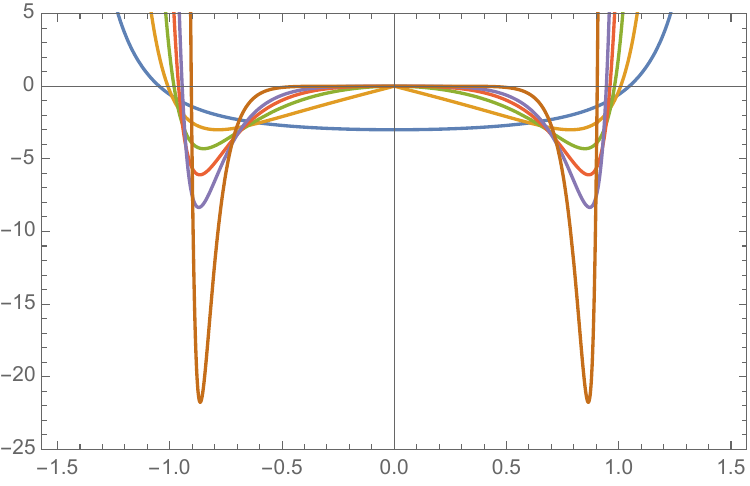}\hfill
\includegraphics[width=0.49\textwidth]{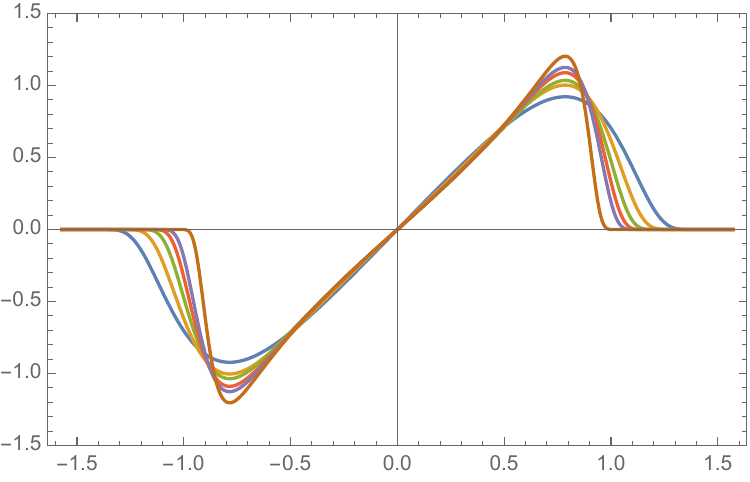}\hfill
\par\end{centering}
\caption{The potential~(\ref{eq:potNM}) for $M=0,1$, $N=2,3,4,5,6,10$ (left) along with the corresponding $E=0$ eigenfunctions (right). Potentials and wave functions are plotted as functions of $y=\arctan x$. The normalization of the wave functions has been chosen such that $\int_{-\pi/2}^{\pi/2} dy\,{\Psi_M^N}^2(x(y))=1$. The potentials become deeper and the corresponding eigenfunctions approach a rectangular ($M=0$) and a see-saw shape ($M=1$) with increasing $N$. Note that $N=2$ is just the usual harmonic oscillator (shifted in energy).}\label{fig:NM0}
\end{figure}

\begin{figure}[h!]
\begin{centering}
\includegraphics[width=0.49\textwidth]{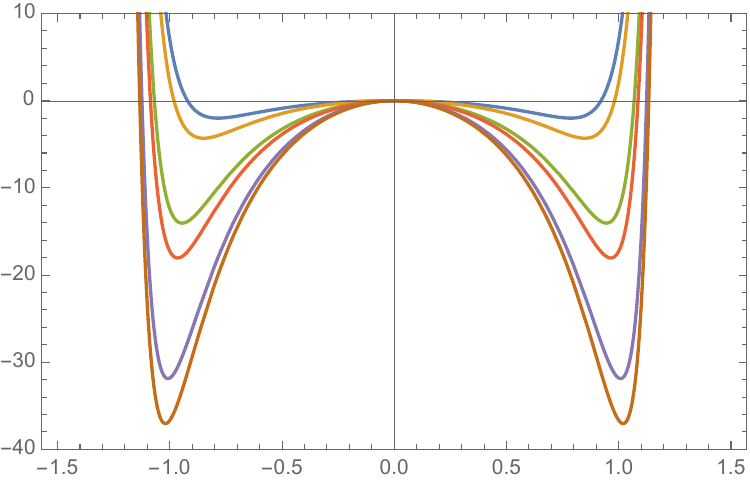}\hfill
\includegraphics[width=0.49\textwidth]{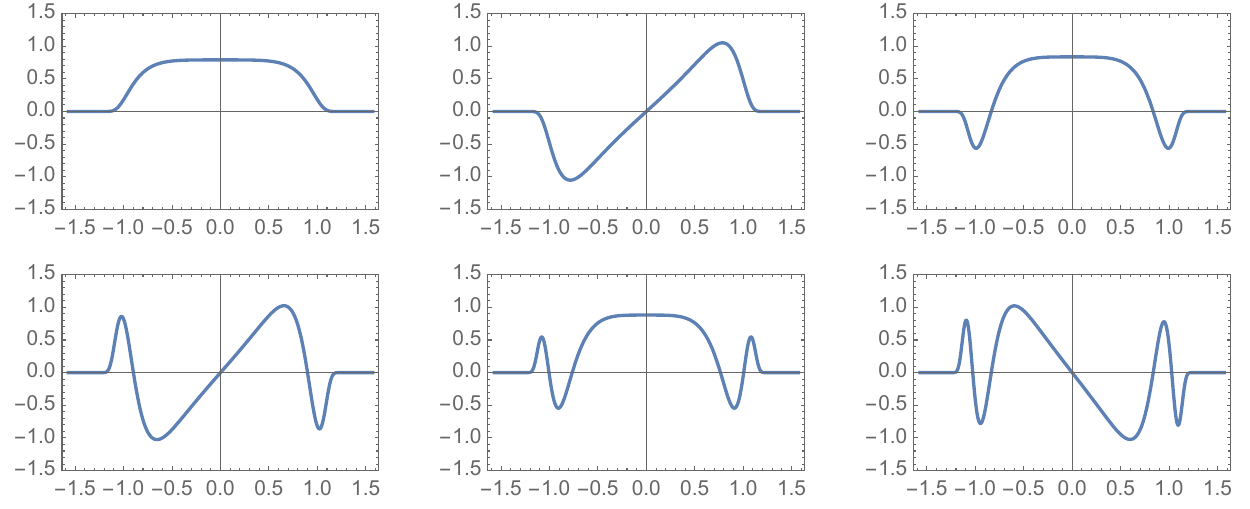}\hfill
\par\end{centering}
\caption{The potential~(\ref{eq:potNM}) for N=4 (sextic potential), M=0,1,4,5,8,9 (left) along with the corresponding $E=0$ eigenfunctions (right). Potentials and wave functions are plotted as functions of $y=\arctan x$. The normalization of the wave functions has been chosen such that $\int_{-\pi/2}^{\pi/2} dy\,{\Psi_M^{\mathrm{sext}}}^2(x(y))=1$. The potentials become deeper with increasing $M$ and the corresponding eigenfunctions exhibit an increasing number of nodes.}\label{fig:4M}
\end{figure}

\medskip
\noindent\underline{$E=0$ solutions for arbitrary $N$ and $M=0,1$:}\\
Potentials of the form (\ref{eq:potNM}) for $M=0,1$ and $N\geq 2$ along with the corresponding $E=0$ eigenfunctions are plotted in Fig.~\ref{fig:NM0} . For $N=2$ it is just the usual harmonic oscillator (shifted in energy). For higher values of $N$ one obtains double-well potentials with increasing depth. For $M=0$ it is the ground state which lies at $E=0$, whereas it is the first excited state for $M=1$. This explains also why, for fixed $N$, the $M=1$ potential is deeper than the $M=0$ potential. The zero-energy eigenfunctions for $M=0$ and $M=1$ are of the form
\begin{equation}
\Psi_0^N(x)= a_0\,e^{-|x|^N/N}\quad\hbox{and}\quad \Psi_1^N(x)= a_1 x \,e^{-|x|^N/N}\, ,\quad N\geq 2\, .
\end{equation}
With increasing $N$, these wave functions seem to approach a rectangular and a  see-saw shape, respectively.

\medskip
\noindent\underline{$E=0$ solutions for $N=4$ and $M=0,1,4,5,8,9$:}\\
Sextic potentials ($N=4$) of the form (\ref{eq:potNM}) with $M$ assuming the six lowest allowed values, are plotted in Fig.~\ref{fig:4M} (left). These are double-well potentials with increasing depth. The corresponding $E=0$ eigenfunctions are shown in Fig.~\ref{fig:4M} (right). They are given by
\begin{equation}
\Psi_M^\mathrm{sext}(x)=\left(\sum_{m=0}^{M} a_{m} x^{m}\right)\,e^{-|x|^4/4}\, ,\quad M=k N, k N+1\, ,\,\, k\in \mathbb{N}_0\, ,
\end{equation} 
where the non-vanishing coefficients $a_m$ are calculated by means of Eq.~(\ref{eq:recursspec}) via downward recursion, starting with $a_M\neq 0$.
For $M=0$ the $E=0$ eigenfunction corresponds to the parity-even ground state. With increasing $M$ the potential becomes deeper, the energy levels go down and at the allowed values of $M$ the excited levels consecutively cross $E=0$. This explains the increasing number of nodes and the pattern of alternating parity-even and parity-odd eigenfunctions. An analogous behavior has already been observed for the quartic potential~\cite{Skala97,Klink:2020akw}.

%
%
%
%
 \section{The electromagnetic field related to polynomial potentials}\label{sec5}
In the foregoing section we have seen that quasi-exactly solvable one-dimensi\-onal Schr\"odinger operators with polynomial interactions can be related to irreducible representations of the nilpotent groups $\mathcal{G}_N$, if the interaction is a particular combination of group generators (see Eq.~(\ref{eq:Hamilton})). Further quasi-exactly solvable problems of physical relevance are obtained by going over to reducible representations. By inducing with the Abelian subgroup 
$ (0, b_1,b_2,\dots,b_{N-1},0)\rightarrow \exp\left(-i \sum_{k=1}^{N-1} \beta_k b_k\right)$  one ends up with a reducible representation of the group $\mathcal{G}_N$. The corresponding generators are then given by
\begin{eqnarray}
(a,0,0,\dots,0) &\rightarrow& X_0 = i\frac{\partial}{\partial x},\nonumber\\
(0, b_1,0,\dots,0) &\rightarrow& X_1= \beta_1 ,\nonumber\\
(0, 0,b_2,0,\dots,0) \, &\rightarrow& X_2 = \beta_2 + \beta_1 x,\nonumber\\
\hspace{2.7cm}&\vdots&\\
(0,\dots,b_{N-1},0) \, &\rightarrow& X_{N-1} = \beta_{N-1} +\dots+\frac{\beta_1 x^{N-2}}{(N-2)!},\nonumber\\
(0,0,\dots, 0,b_N) &\rightarrow& X_N  =  i\frac{\partial}{\partial y}+\beta_{N-1} x+\dots+\frac{\beta_1 x^{N-1}}{(N-1)!}.\nonumber
\end{eqnarray}
With these generators one can now construct a new Hamitonian
\begin{eqnarray}\
\hat{H}^{(\beta_1,\beta_2,\dots,\beta_{N-1})}_\alpha &:=& \,X_0^2 + X_N^2 +\alpha X_{N-1}\\
&=&-\frac{\partial^2}{\partial x^2} +\left(i\frac{\partial}{\partial y}+\beta_{N-1} x+\dots+\frac{\beta_1 x^{N-1}}{(N-1)!} \right)^2
\nonumber\\ &&+\alpha \left(\beta_{N-1} +\beta_{N-2} x +\dots+\frac{\beta_1 x^{N-2}}{(N-2)!} \right)\, ,\nonumber
\end{eqnarray}
in analogy to Eq.~(\ref{eq:Hamilton}). By adding a kinetic term $P_z^2$ for a particle which moves freely in $z$ direction, one ends up with the Hamiltonian
\begin{eqnarray}\label{eq:Hem}
H_\mathrm{em}&:=&\hat{H}^{(\beta_1,\beta_2,\dots,\beta_{N-1})}_\alpha \otimes {I}_z\, \oplus\, {I}_{x\,y}\otimes {P}_z^2 \nonumber\\ 
&=&-\frac{\partial^2}{\partial x^2} +\left(-i\frac{\partial}{\partial y}-\beta_{N-1} x-\dots-\frac{\beta_1 x^{N-1}}{(N-1)!} \right)^2-\frac{\partial^2}{\partial z^2}
\nonumber\\ &&+\alpha \left(\beta_{N-1} +\beta_{N-2} x +\dots+\frac{\beta_1 x^{N-2}}{(N-2)!} \right)\, ,
\end{eqnarray}
where $I_{xy}$ and $I_z$ are unity operators acting on the $(x,y)$ and $z$ coordinates, respectively. But this is just the Hamiltonian for a charged particle moving in the electromagnetic field
\begin{equation}\label{eq:emfield}
\vec{E}(\vec{r})=\left(\begin{array}{c} -\alpha X_{N-2}\\0\\0 \end{array}\right), \qquad
\vec{B}(\vec{r})=\left(\begin{array}{c} 0\\0\\ X_{N-1} \end{array}\right)\, .
\end{equation}
The corresponding electrodynamical potential is
\begin{equation}
\left(A^\mu(x) \right)=\left(\alpha X_{N-1},0,\beta_{N-1} x+\frac{\beta_{N-2} x^{N-2}}{2!} +\dots+\frac{\beta_1 x^{N-1}}{(N-1)!} ,0\right)\, .
\end{equation}
For $N=2$, $H_\mathrm{em}$ reduces to the Hamiltonian for a charged particle moving in a constant magnetic field. It is well known that the corresponding eigenfunctions can be expressed in terms of harmonic-oscillator eigenfunctions~\cite{Landau81}. For $N>2$ the particle moves rather in an $x$-dependent electric and magnetic field. These are perpendicular to each other and the $x$-dependence of the electric field is proportional to the $x$-derivative of the magnetic field. It is also possible in this more general case to relate the energy eigenfunctions $\Phi_\mathcal{E}(x,y,z)$ of $H_\mathrm{em}$ to solutions $\psi_E$ of the eigenvalue problem~(\ref{eq:evproblem}). By means of a Fourier transformation in the $y$ and $z$ variables
\begin{equation}
\Phi_\mathcal{E}(x,y,z)=\frac{1}{2\pi}\int dy\,dz\,e^{ip_y y+ip_z z}\,\tilde\Phi_\mathcal{E}(x,p_y,p_z)
\end{equation}
the eigenvalue problem
\begin{equation}\label{eq:evpem}
H_\mathrm{em}\,\Phi_\mathcal{E}(x,y,z)=\mathcal{E}\,\Phi_\mathcal{E}(x,y,z)
\end{equation}
can be converted into an ordinary differential equation in the $x$-variable
\begin{eqnarray}
\left[-\frac{\partial^2}{\partial x^2} +\left(p_y-\beta_{N-1} x-\dots-\frac{\beta_1 x^{N-1}}{(N-1)!} \right)^2 +p_z^2
+\alpha X_{N-1} \right]&&\hspace{-0.2cm} \tilde\Phi_\mathcal{E}(x,p_y,p_z)\nonumber\\ &&\hspace{-2.0cm}= \mathcal{E}\, \tilde\Phi_\mathcal{E}(x,p_y,p_z)\, ,
\end{eqnarray}
which agrees with Eq.~(\ref{eq:evproblem}), if one makes the following identifications
\begin{equation}
E=\mathcal{E}-p_z^2\qquad\hbox{and}\qquad \beta_N=-p_y\, .
\end{equation}
This means that the Fourier transformation $\tilde{\Phi}_\mathcal{E}(x,p_y,p_z)$ in the $y$ and $z$ variables of a solution $\Phi_\mathcal{E}(x,y,z)$ of the electromagnetic field eigenvalue problem~(\ref{eq:evpem}) gives rise to a solution $\psi_E(x)$ of the polynonial potential problem (\ref{eq:evproblem}) by setting
\begin{equation}
\psi_E(x)=\tilde{\Phi}_\mathcal{E}(x,p_y,p_z)\,\quad\hbox{with}\,\quad \beta_N=-p_y\, ,\,\,E=\mathcal{E}-p_z^2\,\,\hbox{and}\,\,p_y\, , p_z \,\,\hbox{fixed}\, .
\end{equation} 
Note that each $p_y$ is associated with a different value of $\beta_N$ in the corresponding polynomial potential problem. 

More formally, the connection between the electromagnetic field problem and the polynomial potential problem is based on the fact that both Hamiltonians consist of the same combination of infinitesimal generators of the nilpotent group $\mathcal{G}_N$, but in different representations, a reducible and an irreducible one. As a consequence, the Hamiltonian of the electromagnetic field problem can be expressed as a direct integral of Hamiltonians ${H}^{\vec{\beta}}_\alpha$ for the (one-dimensional) polynomial potential problem, i.e.
\begin{eqnarray}
{H}_{\mathrm{em}}&=&\hat{H}^{(\beta_1,\beta_2,\dots,\beta_{N-1})}_\alpha\otimes {I}_z\, \oplus\, {I}_{x\,y}\otimes {P}_z^2\nonumber\\
&=&\left(\int_{\mathbb{R}}^\oplus dp_y\,{H}^{(\beta_1,\beta_2,\dots,\beta_{N-1},\beta_N=-p_y)}_\alpha\right)\otimes {I}_z\, \oplus\, {I}_{x\,y}\otimes {P}_z^2\, .
\end{eqnarray}
Correspondingly the eigenfunctions of the electromagnetic field problem can be decomposed into eigenfunctions of the polynomial potential problem
\begin{eqnarray}\label{eq:efdecomp}
\Phi_\mathcal{E}(x,y,z)&=&\frac{1}{2\pi} \int dp_y\,dp_z\,  e^{i p_y y+i p_z z} \tilde{\Phi}_\mathcal{E}(x,p_y,p_z)\, ,\nonumber\\
&=& \frac{1}{2\pi} \int dp_y\,dp_z\,  e^{i p_y y+i p_z z} {\psi}_{E}(x)\, \delta(E-\mathcal{E}+p_z^2)\, ,\nonumber\\
\end{eqnarray}
with $\beta_N=-p_y$.
Note that $E=E(\alpha,\beta_1,\beta_2,\dots,\beta_{N-1},\beta_N=-p_y)$ is a function of the integration variable $p_y$. Equation~(\ref{eq:efdecomp}) shows, how eigenfunctions and eigenvalues of the one-dimensional polynomial potential problem and a corresponding three-dimensional electromagnetic field problem are related, provided that the boundary conditions in $x$ direction are the same. Since $[{H}_{\mathrm{em}},{P}_y]=[{H}_{\mathrm{em}},{P}_z]=0$, one can look for simultaneous eigenfunctions of $H_{\mathrm{em}}$, ${P}_y$ and ${P}_z$. These are then obviously of the form (see Eq.~(\ref{eq:efdecomp}))
\begin{equation}\label{eq:pwave}
\Phi_{\mathcal{E} p_y p_z}(x,y,z)=\frac{1}{2\pi} e^{i p_y y+i p_z z} \psi_E(x)\,\, \hbox{with}\,\, \beta_N=-p_y\, \hbox{and}\,E=\mathcal{E}-p_z^2\, .
\end{equation}
Here the plane waves have been normalized to a pure delta function.

%
%


\section{Summary}\label{sec6}
In this paper we have looked for quasi-exactly solvable potential models of polynomial form for which part of the spectrum and corresponding eigenfunctions can be calculated analytically. We have been able to identify a large class of such potentials under the assumption that the (one-dimensional) Hamiltonian has the general structure $H_{N}=X_0^2+X_N^2+\alpha X_{N-1}$ and the energy eigenfunctions are of the form $p(x)\,\exp(-\int dx\,X_N)$ with $p(x)$ a polynomial in $X_2$ of degree $M\in \mathbb{N}_0$. Thereby, the $X_i$, $i=0,1,\dots,N$, are the generators of an irreducible representation of a nilpotent group $\mathcal{G}_N$, which can be considered as a generalization of the Heisenberg group $\mathcal{G}_2$ that underlies the solvability of the harmonic oscillator. By inserting the ansatz $p(x)\,\exp(-\int dx\,X_N)$ into the Schr\"odinger equation, we have derived the overdetermined system of equations~(\ref{eq:recursgen}) for the, a priori, unknown polynomial coefficients in $p(x)$. This system of equations is more or less the central result of the paper as far as it holds for polynomial potentials of arbitrary degree $2N-2$ and arbitrary degree $M$ of the polynomial $p(x)$ in the solution ansatz. The coefficients of $p(x)$, as functions of the energy eigenvalue  and the Casimir invariants, follow from $M$ of these equations. The supernumerous equations put constraints on the Casimir invariants and have been used to determine the parameter $\alpha$ as well as the energy eigenvalues. In this way one ends with a three-parameter family of quasi-exactly solvable polynomial potentials of degree $(2N-2)$. This procedure works for even $N\geq 2$. For odd $N$ the ansatz $p(x)\,\exp(-\int dx\,X_N)$ has the wrong asymptotics in the limit $x\rightarrow -\infty$. But what one can do is to symmetrize the potential $X_N^2+\alpha X_{N-1}$ and consider it as function of $|x|$ rather than $x$.  Then $p(x)\,\exp(-\int dx\,X_N)$ is still a solution of the Schr\"odinger equation for $x>0$ and has to be continued as an even or odd function to $x<0$ to end up with a normalizable energy eigenfunction of definite parity for the symmetrized problem. Continuous differentiability of this eigenfunction at $x=0$, however, poses a further constraint on the potential parameters so that one is left with just two free parameters in the symmetrized case. 

In this way we were able to provide a unified treatment of a large class of quasi-exactly solvable polynomial interactions of degree $2N-2$, $N\geq 2$, including the well known, exactly solvable, harmonic oscillator. We have given a lot of examples which show that our approach provides comparably simple expressions for eigenenergies and eigenfunctions, even for $M>2$ (which means higher excitations). Our results for the quasi-exactly solvable (unsymmetrized) sextic ($N=4$) and decatic ($N=6$) potentials extend those known from the literature~\cite{TURBINER1987181,Saad_2006,Turbiner16,MAIZ2018101,Amore:2020pxg,manimegalai2020,Li:2023eji} and \cite{Brandon2013,MAIZ2018101,manimegalai2020}, respectively. Whereas the class of symmetrized quartic potentials provided by our approach~\cite{Klink:2020akw} generalizes the results obtained with the Bethe-ansatz method~\cite{Quesne17}, this is not the case for symmetrized sextic potentials. Our approach and the Bethe-ansatz method lead, in general, to different classes of quasi-exactly solvable potentials which, however, can overlap. In the overlap region we found agreement with the analytic expressions for eigenenergies and wave functions given in Ref.~\cite{Quesne17}. For symmetrized octic potentials ($N=5$) our results seem to be completely new.  New are also the explicit expressions for $E=0$ solutions in the very general case with potential $V_{N,M}(x)=x^{2N-2}-(2M+N-1)\, |x|^{N-2}$, where $N\geq 2$ and $M\geq 0$ can, in principle, be arbitrarily large with some restrictions for possible values of $M$.

Since our algebraization procedure for polynomial interactions in one space dimension was based on irreducible representations of the nilpotent group $\mathcal{G}_N$, it was near at hand to look for new quasi-exactly solvable models by going over to reducible representations. This lead us to the problem of a charged particle in an $x$ dependent magnetic and electric field of polynomial form which are perpendicular to each other. We have shown, that solutions of such problems can be related to one-dimensional polynomial interactions, like the constant magnetic field problem can be related to the harmonic oscillator. This is not the only physics application of the analytic solutions obtained in this paper. Although one has only limited knowledge on the spectrum and the eigenfunctions of quasi-exactly solvable models, they can be of quite some use as starting point and testing ground for approximation schemes and numerical methods.

\section*{Acknowledgements}
The authors acknowledge the financial support by the University of Graz. One of us (W.S.) would like to thank F. Gesztesy for enlightening discussions concerning the domain of the (maximally defined) self-adjoint Schr\"odinger operators associated with the differential expressions (\ref{eq:Hamilton}) and (\ref{eq:hamiltonsym}).

\vfill\break

\end{document}